\documentstyle[12pt,preprint]{aastex}  

\begin{document}

\title{The Colors of Extreme Outer Solar System Objects}
\author{Scott S. Sheppard}    
\affil{Department of  Terrestrial Magnetism, Carnegie Institution of Washington, \\
5241 Broad Branch Rd. NW, Washington, DC 20015 \\ sheppard@dtm.ciw.edu}



\begin{abstract}  

Extreme outer Solar System objects have possible origins beyond the
Kuiper Belt edge, high inclinations, very large semi-major axes or
large perihelion distances.  Thirty-three such objects were observed
in this work to determine their optical colors.  All three objects
that have been dynamically linked to the inner Oort cloud by various
authors (Sedna, 2006 SQ$_{372}$, and 2000 OO$_{67}$) were found to
have ultra-red surface material (spectral gradient, $S \sim 25$).
Ultra-red material is generally associated with rich organics and the
low inclination ``cold'' classical Kuiper Belt objects.  The
observations detailed here show very red material may be a more
general feature for objects kept far from the Sun.  The recently
discovered retrograde outer Solar System objects (2008 KV$_{42}$ and
2008 YB$_3$) and the high inclination object (127546) 2002 XU$_{93}$
show only moderately red surfaces ($S\sim 9$) very similar to known
comets, suspected dead comets, Jupiter and Neptune Trojans, irregular
satellites, D-type asteroids and damocloids.  The extended or detached
disk objects, which have large perihelion distances and are thus
considered to be detached from the influence of the giant planets but
yet have large eccentricities, are found to have mostly moderately red
colors ($10 \lesssim S \lesssim 18$).  The colors of the detached disk
objects, including the dynamically unusual 2004 XR$_{190}$ and 2000
CR$_{105}$, are similar to the scattered disk and Plutino populations.
Thus the detached disk, scattered disk, Plutino and high inclination
``hot'' classical objects likely have a similar mix of objects from
the same source regions.  Outer classical Kuiper belt objects,
including 1995 TL$_{8}$, were found to have very red surfaces ($18
\lesssim S \lesssim 30$).  The low inclination ``cold'' classical
Kuiper Belt objects, outer classical Kuiper belt objects and possibly
the inner Oort cloud appear to be dominated by ultra-red objects ($S
\gtrsim 25$) and thus don't likely have a similar mix of objects as
the other outer Solar System reservoirs such as the scattered disk,
detached disk and Trojan populations.  A possible trend was found for
the detached disk and outer classical Kuiper belt in that objects with
smaller eccentricities have redder surfaces irrespective of
inclinations or perihelion distances.  There is also a clear trend
that objects more distant appear redder.

\end{abstract}


\section{Introduction}

\begin{figure}
\epsscale{0.4}
\centerline{\includegraphics[angle=90,totalheight=0.6\textheight]{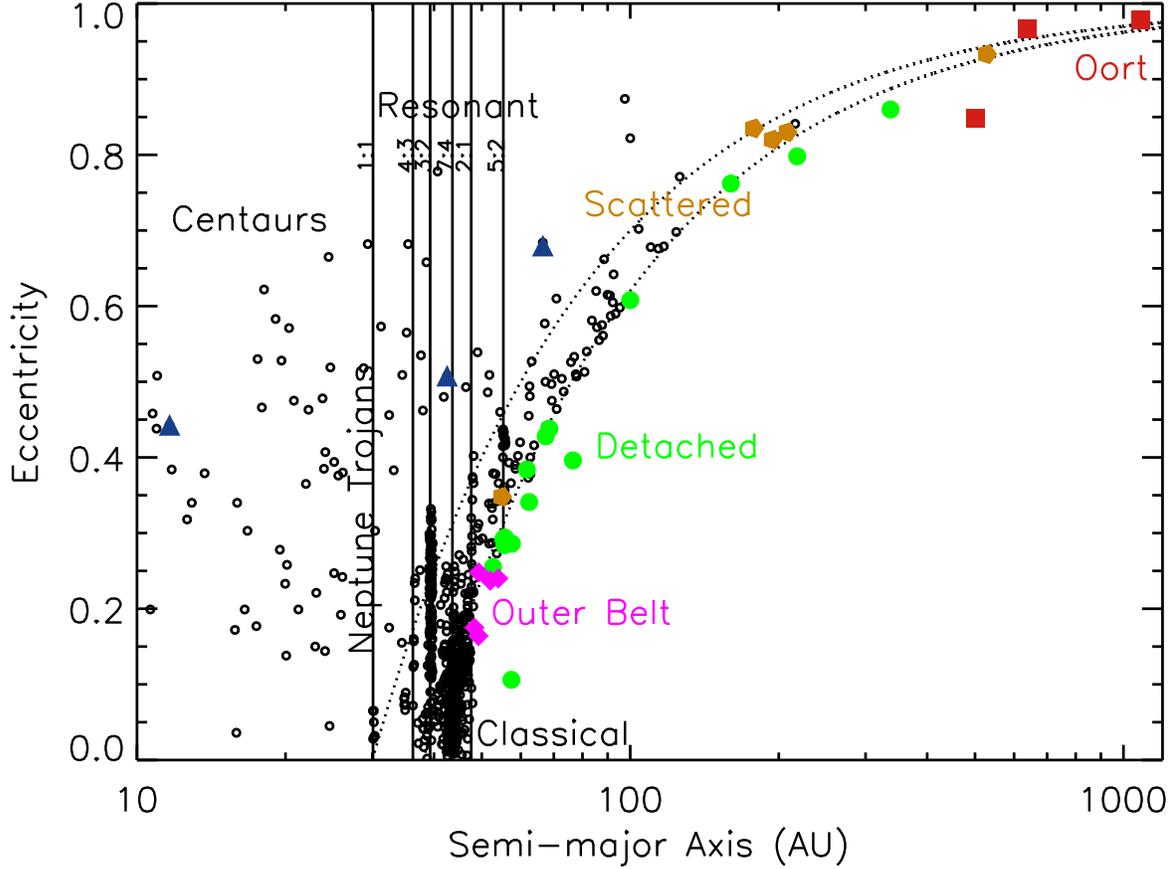}}
\caption{Semi-major axis versus eccentricity of multi-opposition
observed trans-Neptunian objects.  This figure shows several distinct
dynamical KBO populations. Vertical solid lines show resonances with
Neptune as well as the Neptune Trojans in the $1:1$ resonance.
Scattered disk objects have perihelia $30 \lesssim q \lesssim 38$ AU
which are shown by dashed lines.  Classical objects are in the lower
center portion of the figure.  An edge around 48 AU can clearly be
seen for low eccentricity objects.  Centaurs are on unstable orbits
between the giant planets.  The objects observed in this work are
shown by filled symbols: Inner Oort Cloud objects (red squares),
Extended or Detached disk objects (green circles), Outer Classical
Belt objects (purple diamonds), Retrograde and high inclination
objects (blue triangles), Extreme Scattered Disk objects (brown
pentagons).}
\label{fig:kboea2009} 
\end{figure}

The dynamical and physical properties of small bodies in our Solar
System offer one of the few constraints on the formation, evolution
and migration of the planets.  The Kuiper Belt has been found to be
dynamically structured with several observed dynamical classes
(Trujillo et al. 2001; Kavelaars et al. 2008,2009) (see
Figure~\ref{fig:kboea2009}).  Classical Kuiper Belt Objects (KBOs)
have semi-major axes $42 \lesssim a \lesssim 48$ AU with moderate
eccentricities ($e \sim 0.1$) and inclinations.  These objects may be
regarded as the population originally predicted for the Kuiper Belt,
but their relatively large eccentricities and inclinations were
unexpected.  The dynamics of the classical KBOs have shown that the
outer Solar System has been highly modified through the evolution of
the Solar System (Duncan and Levison 1997; Petit et al. 1999; Ida et
al. 2000; Morbidelli and Levison 2004; Gomes et al. 2005).  Resonant
KBOs are in mean motion resonances with Neptune and generally have
higher eccentricities and inclinations than classical KBOs.  These
objects, which include Pluto and the Plutinos in the 3:2 resonance,
were likely captured into these resonances from the outward migration
of Neptune (Malhotra 1995; Hahn and Malhotra 2005; Levison et
al. 2008).  Scattered disk objects have large eccentricities with
perihelia near the orbit of Neptune ($q \sim 25 - 35$ AU).  The
scattered disk objects are likely to have been scattered into their
current orbits through interactions with Neptune (Duncan and Levison
1997; Duncan 2008a; Gomes et al. 2008).

A new class of outer Solar System object, called the extended or
detached disk (Figure~\ref{fig:kboea2009}), has only recently been
recognized (Gladman et al. 2002; Emel'yanenko et al. 2003; Morbidelli \&
Levison 2004; Allen et al. 2006). To date only a few detached disk
objects are known.  The detached disk objects have large
eccentricities, but unlike the scattered disk objects the detached
disk objects have perihelia $q \gtrsim 38$ AU, which do not appear to
be directly caused by Neptune interactions alone (Gladman and Collin
2006; Levison et al. 2008).  Though unexpected, the discovery of these
detached disk objects have given us a new understanding of our Solar
System's chaotic history.

A few objects have been found that have very large semi-major axes and
eccentricities (Sedna, 2006 SQ$_{372}$, and 2000 OO$_{67}$).  Through
dynamical simulations these objects are best described as coming from
the inner Oort cloud (Brown et al. 2004; Kenyon and Bromley 2004;
Morbidelli and Levison 2004; Kaib et al. 2009).  Two objects have also
been found to have retrograde orbits in the outer Solar System (2008
KV$_{42}$ and 2008 YB$_3$).  These two retrograde objects along with
the very high inclination object 2002 XU$_{93}$ ($i \sim 78$ degrees)
could be from the outer Oort cloud or a possibly yet to be discovered
high inclination source region (Gladman et al. 2009).

\begin{figure}
\epsscale{0.4}
\centerline{\includegraphics[angle=90,totalheight=0.6\textheight]{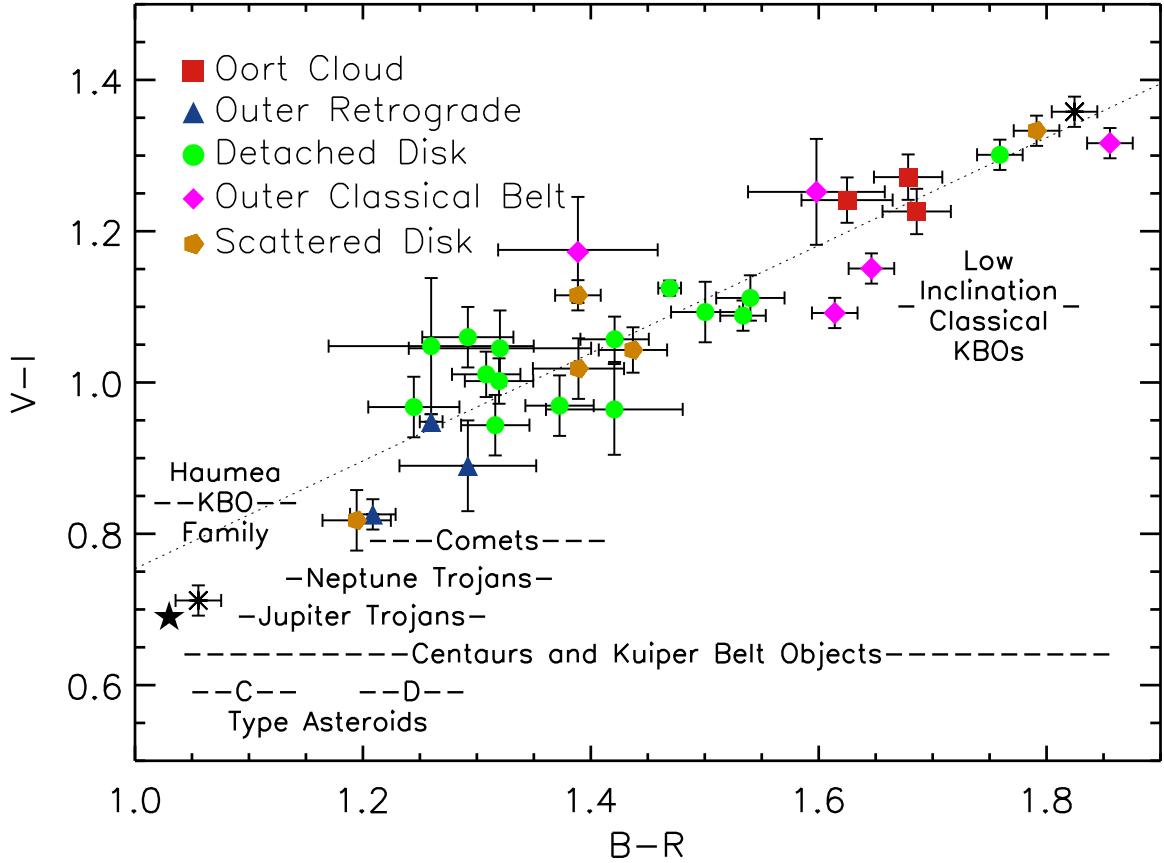}}
\caption{New B-R and V-I colors for objects observed in our sample.
The possible inner Oort cloud objects (red squares) are all near the
ultra-red portion of the figure and are similar to the color of low
inclination Classical Kuiper Belt objects.  The outer retrograde and
high inclination objects (blue triangles) are slightly red and similar
to the colors of the comets, Jupiter Trojans and Neptune Trojans.  The
extended or detached disk objects (green circles) occupy a fairly
large range from moderately red to near but mostly less than
ultra-red.  The outer classical belt objects (purple diamonds) are
mostly near the ultra-red area.  Various extreme scattered disk
objects observed in this work are also shown (brown pentagons).  For
reference the color of the Sun is marked by a filled black star.  The
very neutral colored Haumea collisional family member 1996 TO$_{66}$
and the extremely ultra-red object 1999 OX$_{3}$ (X's) were observed
to show the large range of known colors in the outer Solar System and
confirm the photometry results.  Also shown are the typical B-R colors
found for the C- and D- Type asteroids, Jupiter Trojans, Neptune
Trojans, comets, Haumea collisional family members, low inclination
classical KBOs, Centaurs and Kuiper Belt objects.  The typical colors
of all these objects are generally at the same level and slope as
shown by the dotted line.  The ultra-red material only seen on some
KBOs and Centaurs are shown in the upper right.  Moderately red
objects like the Trojans, comets and some KBOs and Centaurs can be
seen in the middle left of the figure.  Grey or neutral colored
objects like most main belt asteroids are in the lower left of the
figure.  There is an obvious trend that more distant objects appear to
show redder colors.  A B-R color $\gtrsim 1.6$ or V-I color $\gtrsim
1.2$ magnitudes indicates ultra-red color (based on including the
reddest $90\%$ of the low inclination classical KBOs).}
\label{fig:oortcolorall} 
\end{figure}

Some Trans-Neptunian Objects (TNOs) have likely not experienced
significant thermal evolution since their formation.  The amount of
thermal evolution depends on how close the object formed to the Sun
and how close it has approached the Sun during its lifetime (Meech et
al. 2009).  The objects in the Kuiper belt dynamical classes had
varied histories with some experiencing very little thermal evolution,
making them some of the most primitive bodies in the Solar System.
Optical observations of TNOs and Centaurs have shown some of these
objects have the reddest material known in the Solar System
(Figure~\ref{fig:oortcolorall}) (Jewitt and Luu 2001; Peixinho et
al. 2004; Doressoundiram et al. 2008; Tegler et al. 2008).  This
ultra-red material is currently thought to be rich in organic material
(Gradie and Veverka 1980; Vilas and Smith 1985; Cruikshank et
al. 2005; de Bergh et al. 2008).  The ultra-red color may be from
Triton tholins and ice tholins, which can be produced by bombarding
simple organic ice mixtures with ionizing radiation (Doressoundiram et
al. 2003; Barucci et al. 2005a; Emery et al. 2007; Barucci et
al. 2008).

Interestingly, short-period comets that are believed to have
originated from the Kuiper Belt don't show this ultra-red material
(Figure~\ref{fig:oortcolorall}) (Jewitt 2002).  The reason is because
comet surfaces have been highly processed from their relatively close
passages to the Sun (Jewitt 2002; Grundy 2009).  This demonstrates
that the surfaces of comets are not reliable for understanding the
original compositions of the comets.  Some Centaurs, which are the
precursors to the short-period comets, do show these ultra-red colors
probably because they have not yet been near the Sun for a long enough
time to have their surfaces highly modified from thermal, sublimation
or evaporation processes.  No long period comets from the Oort cloud
have been sufficiently observed before any significant heating would
have taken place on their surfaces.  Thus we don't have a good
knowledge of what color an Oort cloud comet may have been before it
started to thermally evolve (Meech et al. 2009).

There have been one or possibly two subsets of TNOs that appear to be
dominated by the ultra-red material (Figure~\ref{fig:oortcolorall}).
First are the low inclination ``cold'' classical Kuiper Belt objects
that also have large perihelions (Tegler and Romanishin 2000; Trujillo
and Brown 2002; Doressoundiram et al. 2005; Gulbis et al. 2006;
Fulchignoni et al. 2008; Peixinho et al. 2008).  These objects likely
formed in the more distant Solar System unlike the higher inclination
KBOs, which may have formed closer to the Sun and were transported to
and captured in the Kuiper Belt during the planet migration process
(Levison and Morbidelli 2003; Gomes 2003; Levison et al. 2008).
Sedna, an object well beyond the Kuiper Belt edge at 50 AU (Jewitt et
al. 1998; Allen et al. 2001), also has an ultra-red color and could be
a new class of object, possibly from the inner Oort cloud (Brown et
al. 2004; Morbidelli and Levison 2004; Kenyon and Bromley 2004;
Brasser et al. 2006; Barucci et al. 2005b).  Some previous works
(Tegler and Romanishin 2000; Trujillo and Brown 2002; Doressoundiram
et al. 2005) have noted that objects with larger perihelion distances
tend to have redder surfaces, but but most of the ultra-red objects
observed were in the main classical Kuiper belt.  No systematic survey
of the colors of the large perihelion detached disk population has
been performed to date.

In this work the optical colors were observed for most of the known
detached disk objects, possible inner Oort cloud objects and other
outer Solar System objects that exhibit extreme orbits in terms of
their inclination, semi-major axis or perihelion distance.
Understanding any color trends or correlations, in particular the
ultra-red material, will constrain where these extreme objects may
have formed in the Solar System and thus how they may have ended up on
their current orbits.  This in turn will allow us to determine how the
planets may have migrated and what amount of this ultra-red organic
rich material may have been incorporated into the planets.

\section{Observations and Analysis}

Observations of the outer Solar System objects presented in this work
were obtained with the twin Magellan Baade and Clay 6.5 meter
telescopes at Las Campanas, Chile and the 8.2 meter Subaru telescope
atop Mauna Kea in Hawaii.  Table 1 shows the various observational
circumstances for the 33 objects observed.  The LDSS3 camera on the
Clay telescope was used on the nights of November 2 and 3, 2005; May 7
and 8 2008; January 28, 2009; May 23 and 24, 2009 and August 25 and
26, 2009. LDSS3 is a CCD imager with one STA0500A $4064\times4064$ CCD
and $15\micron$ pixels.  The field of view is about 8.3 arcminutes in
diameter with a scale of $0.189$ arcseconds per pixel.  The IMACS
camera on the Baade telescope was used on the nights of October 19,
2008 and December 3, 2008.  IMACS is a wide-field CCD imager that has
eight $2048\times4096$ pixel CCDs with a pixel scale of $0.20$
arcseconds per pixel.  The eight CCDs are arranged in a box pattern
with four above and four below and about 12 arcsecond gaps between
chips.  Only chip 2, which is just North and West of the camera
center, was used in the IMACS color analysis.  The Suprime-Cam imager
on the Subaru telescope was used on the night of October 15, 2009.
Suprime-Cam is a wide-field CCD imager that has ten $2048\times4096$
pixel CCDs with a pixel scale of $0.20$ arcseconds per pixel (Miyazaki
et al. 2002).  The ten CCDs are arranged in a $5\times2$ box pattern
similar to the IMACS imager.  Only chip 5, which is just West of the
camera center, was used in the Suprime-Cam color analysis.

\begin{figure}
\epsscale{0.4}
\centerline{\includegraphics[angle=90,totalheight=0.6\textheight]{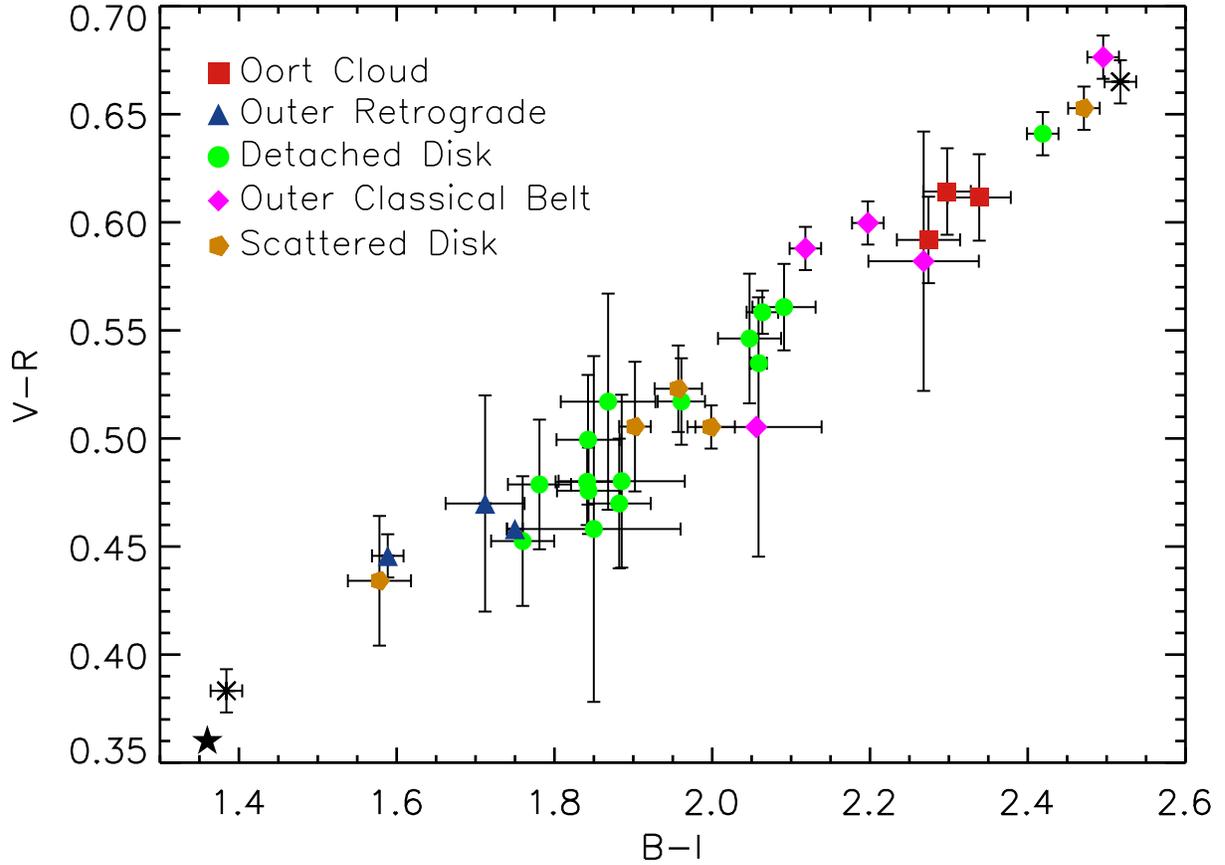}}
\caption{Same as Figure~\ref{fig:oortcolorall} except for B-I and V-R
colors.  Colors with B-I $\gtrsim 2.2$, V-R $\gtrsim 0.6$, or R-I
$\gtrsim 0.6$ magnitudes indicate ultra-red colors (based on including
the reddest $90\%$ of the low inclination classical KBOs).}
\label{fig:oortcolorallBI_VR} 
\end{figure}

\begin{figure}
\epsscale{0.4}
\centerline{\includegraphics[angle=90,totalheight=0.6\textheight]{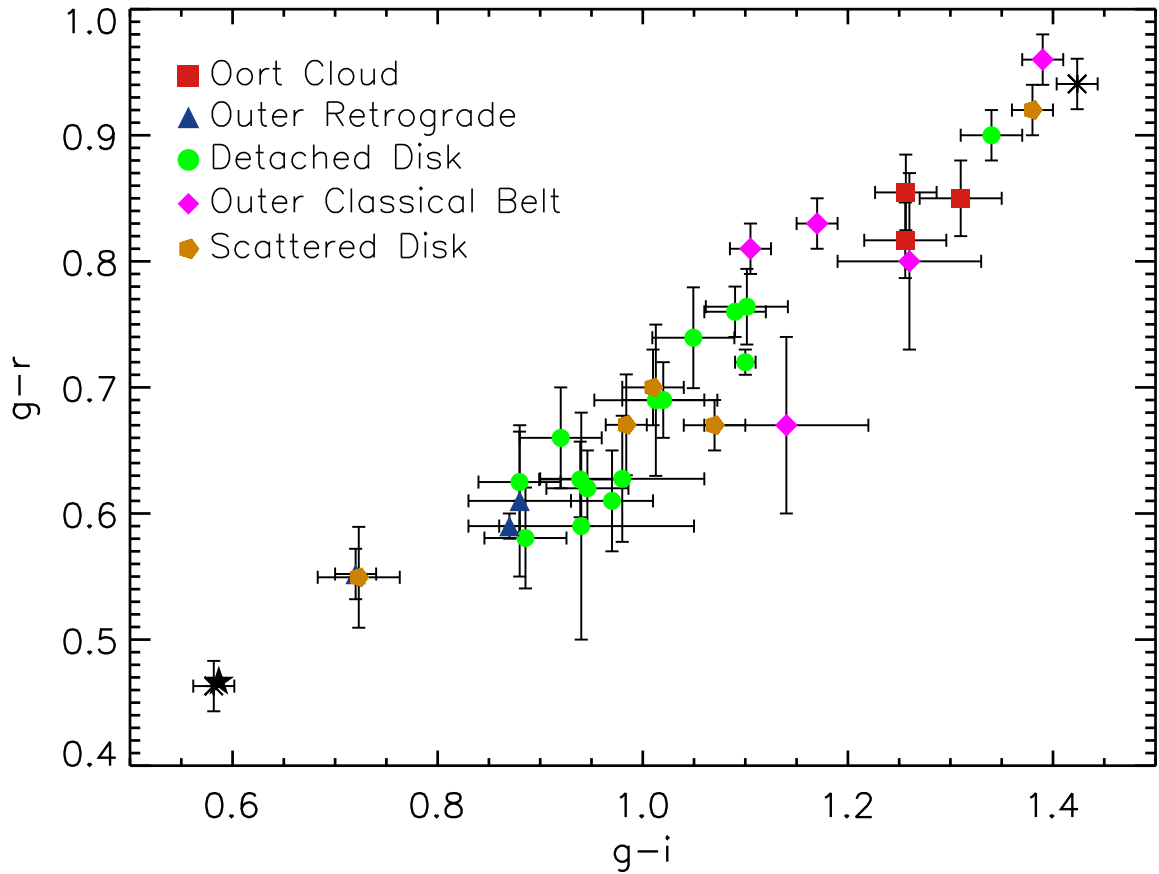}}
\caption{Same as Figure~\ref{fig:oortcolorall} except for Sloan colors
g'-i' and g'-r'.  Colors with g'-i' $\gtrsim 1.2$, g'-r' $\gtrsim
0.8$, or r'-i' $\gtrsim 0.4$ magnitudes indicate ultra-red colors
(based on including the reddest $90\%$ of the low inclination
classical KBOs).}
\label{fig:oortcolorSloan} 
\end{figure}

Dithered twilight flat fields and biases were used to reduce each
image.  Images were acquired through either the Sloan g', r' or i'
filter while the telescope was auto-guiding at sidereal rates using
nearby bright stars.  Exposure times were between 300 and 450 seconds.
Southern Sloan standard stars were used to photometrically calibrate
the data (Smith et al. 2005).  In order to more directly compare our
results with previous works the Sloan colors were converted to the
Johnson-Morgan-Cousins BVRI color system using transfer equations from
Smith et al. (2002).  To verify the color transformation the known
ultra-red (44594) 1999 OX$_{3}$ and grey (19308) 1996 TO$_{66}$ TNOs
were observed (Tegler and Romanishin 1998,2000; Jewitt and Luu 2001;
Barucci et al. 2005a).  The BVRI photometric results are shown in
Table 2 (Figures~\ref{fig:oortcolorall}
to~\ref{fig:oortcolorallBI_VR}) and the Sloan results in Table 3
(Figure~\ref{fig:oortcolorSloan}).

Photometry was performed by optimizing the signal-to-noise ratio of
the faint small outer Solar System objects.  Aperture correction
photometry was done by using a small aperture on the TNOs ($0.\arcsec
57$ to $0.\arcsec 95$ in radius) and both the same small aperture and
a large aperture ($2.\arcsec 46$ to $3. \arcsec 40$ in radius) on
several nearby unsaturated bright field stars with similar Point
Spread Functions (PSFs).  The magnitude within the small aperture used
for the TNOs was corrected by determining the correction from the
small to the large aperture using the PSF of the field stars
(cf. Tegler and Romanishin 2000; Jewitt and Luu 2001).  For a few of
the brighter objects (Sedna, 2003 FY$_{128}$, 2007 JJ$_{43}$, 2008
YB${_3}$) both small apertures and the full large apertures were used
on the TNOs to confirm both techniques obtained similar results.

\section{Results}

The orbital parameters of the 33 outer Solar System objects observed
in this work are shown in Table 4.  There were three main classes of
objects in the observation sample: 1) Objects dynamically linked to
the inner Oort cloud, 2) Outer Solar System retrograde and high
inclination objects and 3) Extended or detached disk and outer
classical belt objects.  Each class is discussed in the subsections
below.  In addition, the well measured grey object (19308) 1996
TO$_{66}$ that is part of the Haumea KBO collisional family and
ultra-red object (44594) 1999 OX$_{3}$ were observed to confirm the
photometry is consistent with previous works.

As seen in Figure~\ref{fig:oortcolorall} all objects observed appear
to have correlated broad band optical colors.  In other words, the
objects appear to follow a nearly linear red slope in the optical.
This has also been confirmed through spectroscopy and correlation
analysis on other TNOs (Doressoundiram et al. 2008).  Because of the
near linearity in the optical colors we can obtain the spectral
gradient, S, of the objects using two unique optical broad band
filters.  The spectral gradient is basically a very low resolution
spectrum of an object and is usually expressed in percent of reddening
per 100 nm in wavelength.  We follow Doressoundiram et al. (2008) and
express the spectral gradient as $S(\lambda_{2} > \lambda_{1}) =
(F_{2,V} - F_{1,V}) /(\lambda_{2} - \lambda_{1})$, where $\lambda_{1}$
and $\lambda_{2}$ are the central wavelengths of the two filters used
for the calculation and $F_{1,V}$ and $F_{2,V}$ are the flux of the
object in the two filters normalized to the V-band filter.  $S$ is the
measure of the reddening of an object's surface determined between two
wavelength measurements (two different filters).  We determined the
spectral gradient of the observed objects using the g' and i' filters,
which have well separated central wavelengths of 481.3 and 773.2 nm
respectively.  The spectral gradient results for the observed objects
are shown in Table 3 and the spectral gradient determined for known
small body populations in the Solar System are shown in Table 5.
Ultra-red color is here defined as including the reddest $90\%$ of the
measured low inclination classical KBOs (Ultra-red: $S\gtrsim 25$, B-R
$\gtrsim 1.6$, V-I $\gtrsim 1.2$, B-I $\gtrsim 2.2$, V-R $\gtrsim
0.6$, R-I $\gtrsim 0.6$, and using Sloan colors g'-i' $\gtrsim 1.2$,
g'-r' $\gtrsim 0.8$, and r'-i' $\gtrsim 0.4$ magnitudes).

\subsection{Inner Oort Cloud Objects}

The Oort cloud is believed to have formed from the scattering of
planetismals from the giant planet region during early planet
formation and is usually separated into two parts (Oort 1950; Stern
2003; Leto 2008; Brasser 2008).  The inner Oort cloud is within a few
thousand to ten thousand AU and is fairly stable to Galactic tides and
passing star perturbations unlike the outer Oort cloud at several tens
of thousands of AU.  While the short period comets are all likely from
the Kuiper Belt region's scattered disk population (Duncan et
al. 2004; Levison et al. 2006), the long period comets are believed to
be from the Oort cloud (Kaib and Quinn 2009).  All the known long
period comets have perihelia within about 10 AU of the Sun.  The
surfaces of the long period comets have already been highly altered
before they are first observed because of the thermal and sublimation
processes that occur as they approach the Sun (Meech et al. 2009).

Sedna was the first object suggested to be part of the inner Oort
cloud (Brown et al. 2004).  Recently Kaib et al. (2009) have suggested
through dynamical simulations that the relatively large perihelia and
semi-major axes of 2006 SQ$_{372}$ and 2000 OO$_{67}$ also make them
likely objects from the inner Oort cloud.  The three inner Oort cloud
objects Sedna, 2006 SQ$_{372}$ and 2000 OO$_{67}$ could be the first
objects from the inner Oort cloud region that we have observed with
thermally unaltered surfaces.  These inner Oort cloud objects are
likely to have formed in a different location than many of the Kuiper
Belt objects.

The observations obtained of these three possible inner Oort cloud
objects in this work show all to be among the reddest objects observed
in this sample.  Their surfaces are, of ultra-red material ($S \gtrsim
25$).  Though all three having ultra-red material is a promising
trend, more inner Oort cloud type objects are needed to be discovered
(see Schwamb et al. 2009) in order to confirm a strong significant ($3
\sigma$) color correlation for inner Oort cloud objects and ultra-red
material.  The spectral gradients of the possible inner Oort cloud
objects are very similar to the red lobe of the Centaur distribution,
the low inclination classical KBOs and outer classical belt KBOs
(Table 5).  As discussed in the introduction, ultra-red material is
likely rich in organic material (Barucci et al. 2008).

\subsection{Retrograde and High Inclination Objects}

Until recently all known retrograde objects had perihelia within the
inner Solar System.  In the last year two objects have been discovered
with retrograde orbits and perihelia in the giant planet region.
Neither shows any current evidence of cometary activity.  2008
YB$_{3}$ has a perihelion around 6.5 AU and thus is likely to have
undergone surface sublimation and interior recrystallization during
its lifetime (Meech et al. 2009).  2008 KV$_{42}$ has a perihelion of
about 21 AU and thus the amount of surface alteration of this object
could be significantly less than other retrograde objects and comet
type objects.  Gladman et al. (2009) simulated the orbit of 2008
KV$_{42}$ and found its perihelion distance likely has not been
interior to Saturn over the age of the Solar System.  It is unknown
where 2008 KV$_{42}$ came from but its orbit is similar to Halley's
comet and thus it could have come from the Oort cloud or another yet
to be discovered high inclination reservoir.

The observations obtained of these two outer Solar System retrograde
objects and the similar high inclination object (127546) 2002
XU$_{93}$ show all to have only moderately red surfaces ($S \sim 9$).
Their spectral gradients are similar to the known comets, extinct
comet objects, Jupiter Trojans, Neptune Trojans, irregular satellites
and damocloids (Table 5).  This suggests the outer retrograde and high
inclination object surfaces have been thermally altered over the age
of the Solar System as is expected for these other similarly
moderately red colored volatile rich objects.

\subsection{Extended/Detached Disk and Outer Classical Belt Objects}

Objects with large semi-major axes and perihelion distances have only
recently been discovered (Gladman et al. 2002).  Knowledge of the
physical properties of these dynamically interesting objects is
important to constrain their origins and evolution.  Detached disk
objects are considered to have moderate to large eccentricities ($e>
0.2 - 0.25$), large perihelion distances ($q \gtrsim 38$ AU) and large
semi-major axes ($50 \lesssim a \lesssim 500$ AU)(Elliot et al. 2005;
Lykawka and Mukai 2007a; Gladman et al. 2008).  Detached disk objects
are somewhat decoupled from the giant planet region yet have been
considerably influenced dynamically to obtain their relatively large
eccentricities.  The objects in the detached disk can thus be
considered intermediate between the Kuiper Belt and the inner Oort
cloud.  Objects with dynamics closely related to the detached disk are
the outer classical belt population.  The outer classical belt objects
have $a>48.4$ AU, $e<0.25$ and are nonresonant (Gladman et al. 2008).
Objects with $39.4<a<48.4$ AU and $e<0.25$ are considered main
classical belt objects or cubewanos.  The $2:1$ Neptune resonance
separates the main classical belt from the outer classical belt.

In this work most of the known detached disk and outer classical belt
objects were observed to determine their optical colors for the first
time in order to compare them to other Solar System small body
reservoirs.  In particular, determining if these populations are
dominated by ultra-red material allows important constraints to be
placed on the origin and evolution of these populations.

\subsubsection{Detached Disk}

Though the detached disk has been defined differently by various
authors this work takes a very strict definition.  A detached disk
object must have $q>38$ AU, $e>0.25$ and $50<a<500$ AU.  Thus 13
objects that were observed in this work qualify as detached disk
objects under this definition (2008 OG$_{19}$, 2005 SD$_{278}$, 2005
TB$_{190}$, 2004 OJ$_{14}$, 2004 VN$_{112}$, 2003 QK$_{91}$, 2003
EO$_{297}$, 2003 FZ$_{129}$, (84522) 2002 TC$_{302}$, 2000 CR$_{105}$,
2000 YW$_{134}$, (118702) 2000 OM$_{67}$, 1999 HW$_{11}$).

The colors of the detached disk objects do not appear to be
extraordinary (Figure~\ref{fig:oortcolorall}).  Except for one
ultra-red detached disk object, the rest show only moderately red
colors ($10 \lesssim S \lesssim 18$).  Their spectral gradient average
($S=14.5\pm5$) is very similar to the scattered disk KBOs, Plutinos,
high inclination classical KBOs as well as the damocloids and comets
(Table 5).  The detached disk objects are thus not likely from the
same source region as the ultra-red low inclination classical KBO
population or the inner Oort cloud though if they are from the same
source region than the detached disk objects had significantly
different surface altering histories.  Inclination is not important in
the color of detached disk objects with even the few very low
inclination objects observed in the detached disk (2003 FZ$_{129}$ and
2003 QK$_{91}$) showing only moderately red colors.  The discovery of
more low inclination detached disk objects are needed to further
confirm that this population is not rich in ultra-red material unlike
the low inclination main classical belt.  The only detached disk
object found to have ultra-red surface material is (84522) 2002
TC$_{302}$, which has a large inclination of 35 degrees.  (84522) 2002
TC$_{302}$ is possibly in the 5:2 Neptune resonance and as discussed
below it appears objects in high order Neptune resonances are on
average very red.

\subsubsection{Outer Classical Belt Objects}
The outer classical belt objects have $a>48.4$ AU, $e<0.25$, $i<40$
degrees and are non-resonant.  Outer classical belt objects are
separated from the main classical belt by the 2:1 resonance and have
slightly smaller eccentricities than the detached disk objects.  The
observed sample has 5 bonafide outer classical belt objects (2007
JJ$_{43}$, 2003 FY$_{128}$, 2003 UY$_{291}$, 2001 QW$_{297}$ and
(48639) 1995 TL$_{8}$).

The only other possible outer disk object in our sample would be 2004
XR$_{190}$.  This is a very dynamically unusual object since it has a
relatively low eccentricity, large semi-major axis and large
inclination (Table 4).  It is to date a dynamically unique object but
has been classified as an outer disk object by Gladman et al. (2008)
and a detached disk object by Allen et al. (2006) and Lykawka and
Mukai (2007a).  Gomes et al. (2008) believe that 2004 XR$_{190}$ is a
fossil detached object.  2004 XR$_{190}$ was likely scattered by a
close planetary encounter into the $3:8$ mean motion resonance with
Neptune.  2004 XR$_{190}$ then escaped from the $3:8$ mean motion
resonance while Neptune was still migrating outwards during the very
early evolution of the Solar System.  Scattering and escaping the mean
motion resonance would help explain the rather large inclination,
large perihelion distance and large size of 2004 XR$_{190}$.  In
addition, the Gomes (2003) model found outer classical belt objects
are not expected to obtain such high inclinations as 2004 XR$_{190}$.
2004 XR$_{190}$ has only a moderately red color of $S=10.3\pm3$ like
the higher eccentricity detached objects (further discussion of 2004
XR$_{190}$ is in section 4).

Excluding the dynamically unique 2004 XR$_{190}$, the outer classical
belt objects are significantly redder ($18 \lesssim S \lesssim 30$)
than the average detached disk objects ($10 \lesssim S \lesssim 18$).
The average outer classical belt objects spectral gradient
($S=22.8\pm5$ or $S=21.0\pm5$ if including 2004 XR$_{190}$) is similar
to the ultra-red material seen in the low inclination classical Kuiper
belt and inner Oort cloud objects (Table 5).  The sample of outer
classical belt objects are all very red even though they cover a wide
range of inclinations with both 2003 UY$_{291}$ and 1995 TL$_{8}$
having very low inclinations ($i<4$ degrees) and 2001 QW$_{297}$, 2003
FY$_{128}$ and 2007 JJ$_{43}$ having moderate inclinations ($i \sim
12-17$ degrees).  This is unlike the main classical belt were the low
inclination objects are dominated by ultra-red objects ($S\gtrsim 25$)
while the higher inclination objects are not dominated by ultra-red
material.  More outer classical belt objects need to be discovered to
confirm this population is dominated by very red objects ($S\gtrsim
20$).

\section{Discussion}

\subsection{Detached and Scattered Disk}

The scattered disk is probably made up of two main source populations.
Some scattered disk objects are likely the surviving members of a
relic population of objects that were scattered during Neptune's
migration in the very early Solar System (Gomes et al. 2008).  A
second source for the scattered disk is from recently dislodged
objects from the Kuiper Belt through various slow dynamical processes
(resonances) or collisions (Duncan et al. 1995; Levison and Duncan
1997; Duncan and Levison 1997; Nesvorny and Roig 2001; Gomes et
al. 2008).

How the detached disk may have formed is still an open question (Gomes
et al. 2008; Morbidelli et al. 2008; Kenyon et al. 2008; Duncan et
al. 2008b; Gladman et al. 2008).  For high inclination objects ($i>50$
degrees) the Kozai resonance can allow scattered objects to obtain
large perihelion distances (Thomas and Morbidelli 1996; Gallardo
2006).  For objects with moderate inclinations the Kozai mechanism
only works in increasing the perihelion distance of a scattered object
if the object is also in a mean motion resonance with Neptune (Gomes
2003).  Using Neptune mean motion resonances and the Kozai mechanism
Gomes et al. (2008) believes the high perihelia and relatively large
semi-major axes of some moderate inclination detached objects can be
explained through the above mechanism, specifically 2000 YW$_{134}$,
2005 EO$_{297}$ and 2005 TB$_{190}$ as well as the high inclination
object 2004 XR$_{190}$, since they are all in or near Neptune mean
motion resonances.  These objects were likely at some point scattered
disk objects that simply had their perihelia raised through Neptune
mean motion resonances and the Kozai effect.  

Based on the similar average spectral gradients of the two populations
the origin of the objects in the detached disk could be similar as the
scattered disk (Table 5).  The scattered disk spectral gradient ($S =
10.1\pm5$) shown in Table 5 uses the strict definition similar to
Gladman et al. (2008) which eliminates objects thought to be in any
resonance with Neptune from being called a scattered disk object
(called here the strict scattered disk: objects not in an obvious high
order resonance with Neptune, perihelia less than 35 AU and semi-major
axis between 30 and 100 AU).  If objects in high order resonances with
Neptune are allowed in the definition used for what is a scattered
disk object the spectral gradient increases slightly and is almost the
same as the detached disk average spectral gradient ($14.5\pm5$).  It
is interesting to note that very red ($S\gtrsim 20$) objects are
absent in the strict definition of the scattered disk but are not when
including the higher order resonance objects.  This may hint that many
high order resonance scattered disk objects are coming from the
ultra-red low inclination classical belt or outer classical belt
objects.  It may be that the only efficient way to dislodge these
fairly dynamically stable ultra-red objects is through some resonance
interactions.

To further compare the scattered disk to the detached disk population
the Student's t-test and the Kolmogorov-Smirnov (K-S) test were
performed on the spectral gradients of the two populations
(Figure~\ref{fig:KStestDetachScatterStrict}).  The differences in the
two population distributions were not statistically significant ($<
3\sigma$) in either test and thus are consistent with both populations
coming from the same parent population (Table 6).  This is true no
matter if the high order outer resonance objects are considered
scattered disk objects or not (Figure~\ref{fig:KStestDetachScatter}).
The similarity of spectral gradients may hint that Neptune mean motion
and Kozai resonances allowed scattered disk objects to become detached
overtime from significant Neptune influence and that the detached disk
is a simple extension of the scattered disk (Gallardo 2006; Lykawka
and Mukai 2007b; Emel'yanenko and Kiseleva 2008; Gladman et al. 2008;
Gomes et al. 2008).  Based on the spectral gradients and dynamics of
the objects in the detached and scattered disk it appears they likely
contain many objects from the same source region.

\subsection{Ultra-red Colors and the Outer Classical Belt}

The outer classical belt objects have lower eccentricities and usually
lower semi-major axes than the detached disk objects.  They are
separated from the main classical belt by the Neptune $2:1$ mean
motion resonance.  The dynamical origin of the outer classical belt
objects are not easy to explain through simple Neptune mean motion
resonances and the Kozai effect and may have a different origin than
the detached disk objects (Gomes 2003; Gomes et al. 2008; Morbidelli
et al. 2008).  Simulations by Gomes (2003) of Neptune's migration and
the formation of the Kuiper belt show that the objects coming from the
outer most portion of the disk that Neptune migrates through would
have preferentially low inclinations ($i<10$ degrees) and low
eccentricities ($e\lesssim 0.1$) when dispersed to near 40 AU.  This
is likely the source of the ``cold'' classical disk (see Gomes (2003)
Figure 2).  The inclination distribution for these objects is found in
the simulations to increase slightly at larger semi-major axes.  More
importantly, the Gomes simulations show that these same objects
further out in semi-major axis around 50 AU would have significantly
larger eccentricities ($e\sim 0.2$).  Using these ideas Gomes et
al. (2008) suggest that objects with orbits like the outer classical
belt are not fossilized detached disk objects and more likely share a
similar origin as the low inclination ``cold'' classical population
(Gomes 2003; Morbidelli et al. 2008).  The very red colors ($S\gtrsim
20$) found in this work for these outer classical belt objects support
this hypothesis.  The spectral gradient of the outer classical belt
objects averages $S=23.3\pm5$ which is similar to that found for the
low inclination ``cold'' classical main belt objects ($27.5\pm5$: see
Table 5).

\begin{figure}
\epsscale{0.4}
\centerline{\includegraphics[angle=90,totalheight=0.6\textheight]{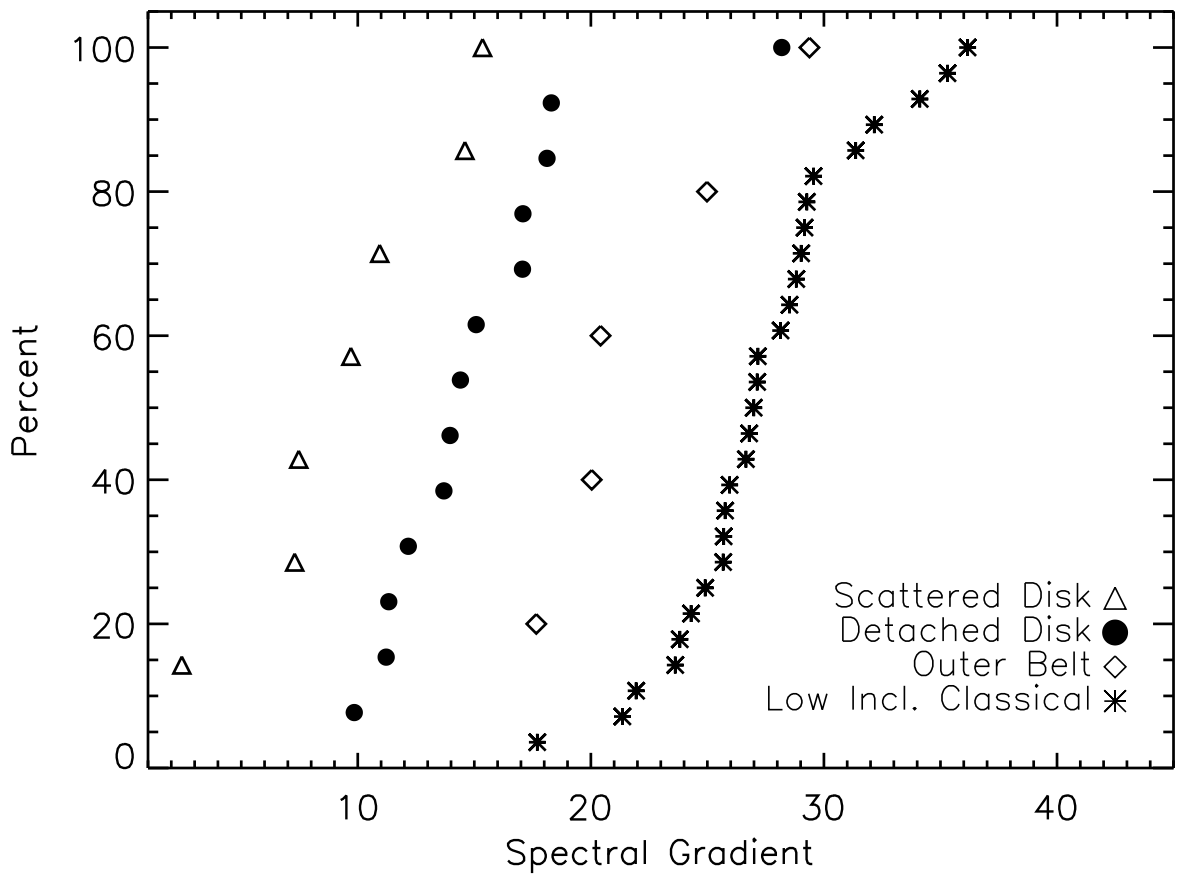}}
\caption{The Kolmogorov-Smirnov test (K-S test) plotted for the
detached disk (circles), outer classical belt (diamonds), low
inclination ``cold'' classical belt (asterisks) and strict scattered
disk objects (triangles: not including objects thought to be in high
order resonances with Neptune, having perihelia above 35 AU or
semi-major axes above 100 AU).  The vertical axis shows the cumulative
spectral gradient for the objects.  It is clear that the groups have
some overlap in color but on average the low inclination classical
belt objects are the reddest followed by the outer classical belt
objects, the detached disk objects and the most neutral objects being
the scattered disk.  The results of comparing various population
spectral gradient distributions are shown in Table 6.}
\label{fig:KStestDetachScatterStrict} 
\end{figure}

\begin{figure}
\epsscale{0.4}
\centerline{\includegraphics[angle=90,totalheight=0.6\textheight]{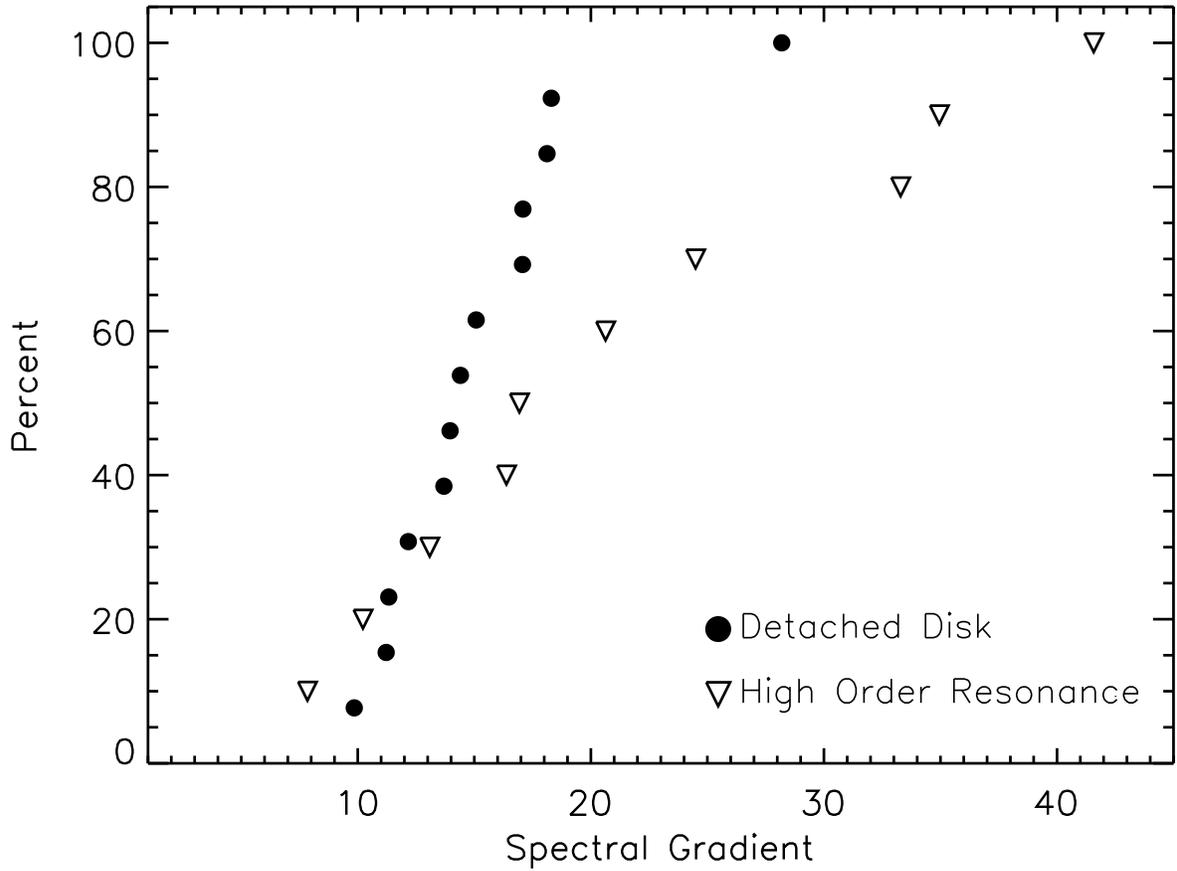}}
\caption{Same as Figure~\ref{fig:KStestDetachScatterStrict} except
showing scattered disk objects thought to be in high order resonances
with Neptune (upside down triangles).  The high order resonance
objects consist of a wide range of spectral gradients including a
significant amount of ultra-red objects.}
\label{fig:KStestDetachScatter} 
\end{figure}

To compare the spectral gradients of outer classical belt objects with
the low inclination ``cold'' classical belt objects the Student's
t-test and Kolmogorov-Smirnov test were performed on the two
populations (Figure~\ref{fig:KStestDetachScatterStrict}).  The two
distributions do not appear to be significantly different ($<2\sigma$)
and thus could come from the same parent population (Table 6).  This
is unlike the detached and strict scattered disk which have $>3\sigma$
confidence in the differences of their spectral gradient distributions
when compared to the low inclination ``cold'' classical belt objects
(Table 6).  Thus the detached disk and strict scattered disk objects
are unlikely to have come from the same parent population as the low
inclination ``cold'' classical belt objects.

Table 6 shows that the K-S test hints at a possible trend with there
being significant differences between the outer classical belt
spectral gradient distribution and the strict scattered and detached
disk objects but with only five known outer classical belt objects the
test is unreliable.  About twice as many outer classical belt objects
need to be discovered and have their spectral gradients determined in
order to confirm or reject them as having significantly different
spectral gradients from the various dynamical populations in the outer
solar system.  It is apparent that the outer classical belt objects
are very red objects and they are redder than both the detached disk
and strict scattered disk and less red than the low inclination
``cold'' classical KBOs.

As shown in Figure~\ref{fig:KStestDetachScatterStrict} the colors of
the scattered disk objects not in resonances are the least red.  The
detached disk objects are slightly redder while the outer classical
belt objects are even redder and finally the low inclination ``cold''
classical KBOs are the reddest objects.  The high order resonance
objects appear to span most of the spectral gradient range of the
various populations (Figure~\ref{fig:KStestDetachScatter}).  The
significant differences in spectral gradients for some of the
populations is likely because the objects come from different source
regions.  It is also possible that the differences in the spectral
gradients of the various populations comes from significantly
different surface weathering processes on the objects over the age of
the Solar System, such as different collisional or sublimation
histories.  It is apparent that the objects more distant from the Sun
are on average redder.

\subsection{Spectral Gradients Versus Orbital Dynamics}

\begin{figure}
\epsscale{0.4}
\centerline{\includegraphics[angle=90,totalheight=0.6\textheight]{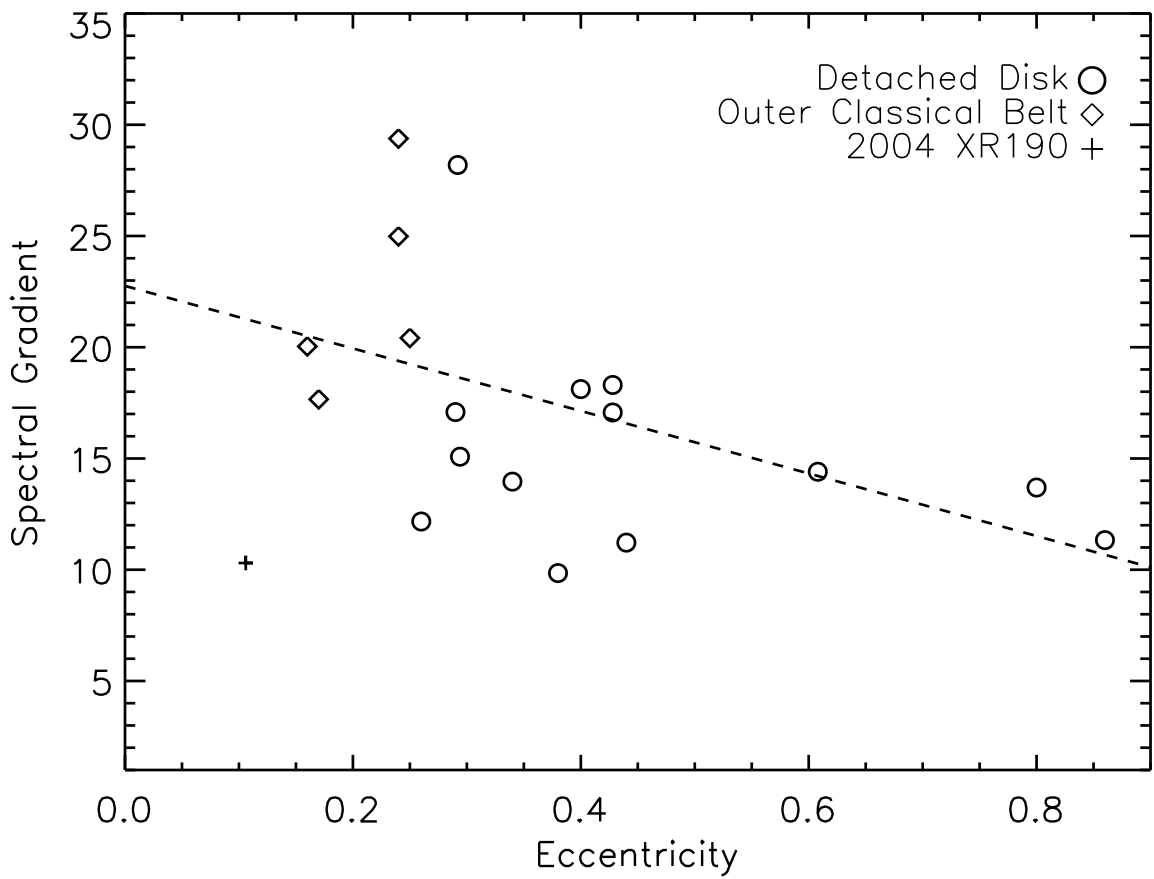}}
\caption{The eccentricity versus the spectral gradient for 2004
XR$_{190}$, detached disk and outer classical belt objects.  There
appears to be a trend that the lower the eccentricity the redder the
object, but since there is only a few objects in the sample this trend
is only at the $97\%$ confidence level using the Pearson correlation
coefficient.  The lower eccentricity outer classical belt objects
(diamonds) are near the ultra-red spectral gradient region while the
higher eccentricity detached disk objects (circles) are mostly
moderately red to neutral in color.  2004 XR$_{190}$ is dynamically
distinct (see text) but has been simulated as a detached disk object
by Gomes et al. (2008) and thus is plotted for completeness (plus
sign).  A linear fit is shown by the dashed line.}
\label{fig:BIvsEccent} 
\end{figure}

\begin{figure}
\epsscale{0.4}
\centerline{\includegraphics[angle=90,totalheight=0.6\textheight]{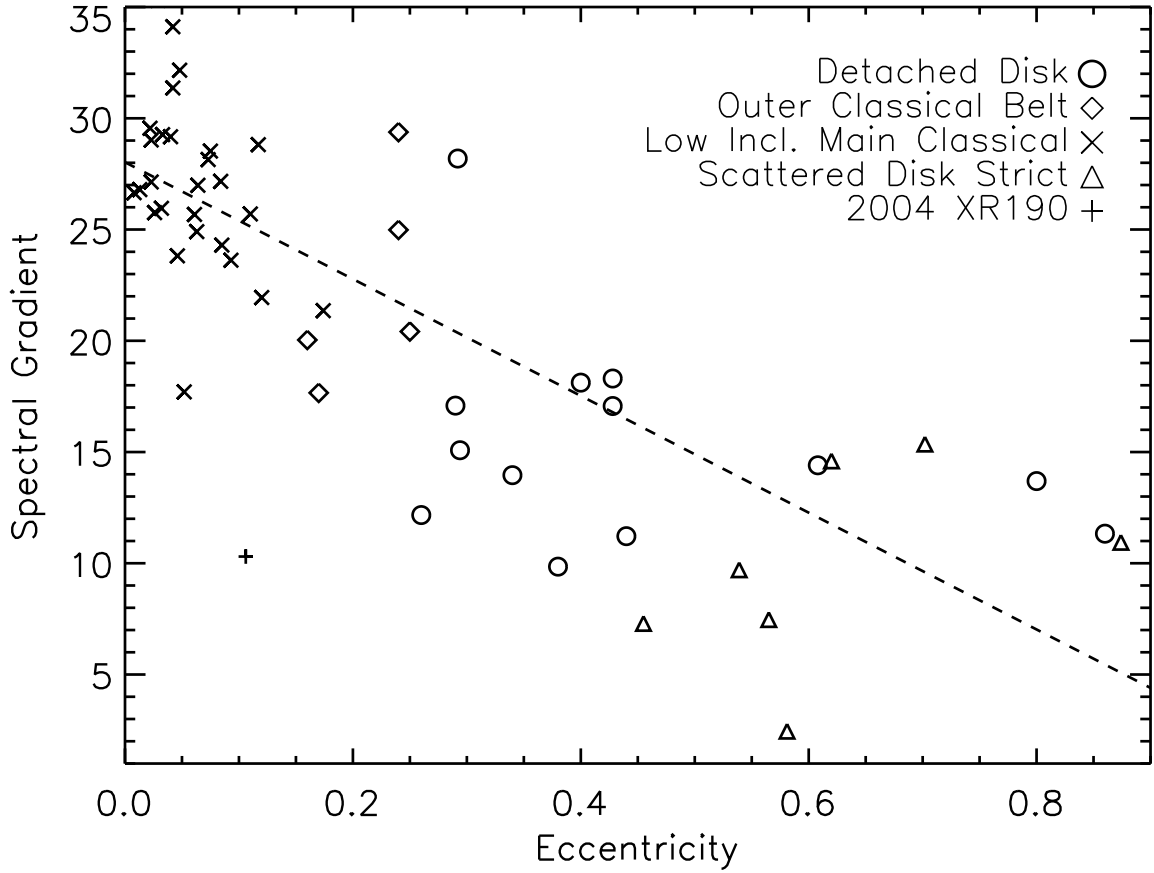}}
\caption{Same as Figure~\ref{fig:BIvsEccent} except now the scattered
disk objects not in high order resonances (triangles) and the low
inclination ``cold'' main classical KBOs (X's) have been added.
Adding these objects strengthens the trend that lower eccentricity
objects have redder colors and is at the $99.99\%$ confidence level.}
\label{fig:BIvsEccentall} 
\end{figure}

To further explore the origins of the detached disk and outer
classical belt objects their eccentricities versus spectral gradient
were plotted (Figure~\ref{fig:BIvsEccent}).  There is an apparent
trend that the lower the eccentricity the redder the object.  The
Pearson correlation coefficient is -0.49 using the eighteen known
spectral gradients of the detached disk and outer classical belt
objects.  The correlation with eccentricity is only significant at
about the $97\%$ level and additional low eccentricity outer classical
belt objects need to be found to confirm or reject this possible trend
(Table 7).  Including the strict scattered disk objects increases the
significance of the correlation with eccentricity to $99.1\%$.  If the
low inclination ``cold'' main classical KBOs are also included the
trend is even stronger with a Pearson correlation coefficient of -0.80
and a significance at the $99.99\%$ confidence limit
(Figure~\ref{fig:BIvsEccentall}).  There is no trend of spectral
gradient with the inclination or perihelion distances of the detached
disk and outer classical belt objects (Table 7).  Including the strict
scattered disk also finds no trend with spectral gradient and
inclination or perihelion distance.

\section{Summary}

Thirty-three extreme outer Solar System objects were observed to
determine their optical colors.

1) The three possible inner Oort cloud objects (Sedna, 2006
SQ$_{372}$, and 2000 OO$_{67}$) all have ultra-red surfaces (spectral
gradient $S \sim 25$).  These ultra-red surfaces are abundant in the
low inclination ``cold'' classical KBO population and is believed to
be associated with organic-rich material.  Because the ultra-red
material is only seen in the very outer parts of the observable Solar
System it is likely this material has not been significantly thermally
altered.  The red lobe of the Centaur distribution could thus either
be from the low inclination classical KBO population or from the inner
Oort cloud population.

2) For the first time a systematic color determination of extended or
detached disk objects was obtained. Most detached disk objects have
only moderately red surfaces ($10 \lesssim S \lesssim 18$).  Though
slightly redder on average than the scattered disk the detached disk
colors are consistent with being from the same source region as the
scattered disk objects.  The only ultra-red objects observed with
scattered disk like orbits appear to be objects in high order
resonances with Neptune.

3) The outer classical Kuiper belt objects, which have semi-major axes
   beyond the $2:1$ resonance with Neptune and low eccentricities,
   were found to be very red ($S \gtrsim 20$) and are on average
   redder than the detached disk objects.  Unlike the scattered disk
   and detached disk the outer classical belt objects have spectral
   gradients similar to the ultra-red low inclination ``cold''
   classical KBOs though they appear to be less red on average.

4) The two retrograde objects with perihelia in the outer Solar System
(2008 KV$_{42}$ and 2008 YB$_3$) and the extremely high inclination
object (127546) 2002 XU$_{93}$ show only moderately red colors ($S
\sim 9$).  These colors are very similar to the known comets, dead
comets, damocloids, Jupiter Trojans, Neptune Trojans, irregular
satellites, D-type main belt asteroids, scattered disk objects and the
neutral lobe of the Centaurs.  2008 YB$_3$ perihelion is near Jupiter
thus this object has had its surface thermally altered over the age
the of the Solar System as is probably true for all the above
moderately red populations.  2008 KV$_{42}$ has a rather large
perihelion at 21 AU and it is unknown if it has ever approached closer
to the Sun.  The moderately red surface color suggests its surface has
likely been thermally altered.

5) The detached disk and outer classical Kuiper belt objects show a
   trend that the lower the eccentricity the redder the object.  This
   trend is currently not statistically significant since only a few
   of these objects are known.  The trend is strengthened when adding
   the strict scattered disk and low inclination ``cold'' classical
   KBOs.  The trend must be confirmed through discovering and
   measuring the colors of more outer classical belt objects.

\section*{Acknowledgments}

I thank Chadwick A. Trujillo and Henry H. Hsieh for helpful comments
that improved this manuscript.  This paper includes data gathered with
the 6.5 meter Magellan Telescopes located at Las Campanas Observatory,
Chile.  Based in part on data collected at Subaru Telescope, which is
operated by the National Astronomical Observatory of Japan.  This work
was partially supported by the National Aeronautics and Space
Administration through the NASA Astrobiology Institute (NAI) under
Cooperative Agreement No. NNA04CC09A issued to the Carnegie
Institution of Washington.


\newpage

%
%
%
%



\begin{center}
\begin{deluxetable}{llccc}
\tablenum{1}
\tablewidth{6 in}
\tablecaption{Geometrical Circumstances of the Observations}
\tablecolumns{5}
\tablehead{
\colhead{Name} & \colhead{UT Date} & \colhead{$R$}  & \colhead{$\Delta$} & \colhead{$\alpha$} \\ \colhead{} &\colhead{} &\colhead{(AU)} &\colhead{(AU)} &\colhead{(deg)} }  
\startdata
(90377) Sedna                        & 2008 Dec 03.267-.291  & 87.95 & 87.05 & 0.25  \nl   
(48639) 1995 TL$_{8}$                & 2009 Aug 25.289-.347  & 43.86 & 43.52 & 1.25  \nl
                                     & 2009 Aug 26.359-.373  & 43.86 & 43.50 & 1.24  \nl
(19308) 1996 TO$_{66}$               & 2005 Nov 02.017-.028  & 46.47 & 45.64 & 0.67  \nl   
(181874) 1999 HW$_{11}$              & 2008 May 07.234-.299  & 42.99 & 42.00 & 0.26  \nl  
(44594) 1999 OX$_{3}$                & 2005 Nov 03.021-.033  & 24.48 & 24.36 & 2.31  \nl   
(148209) 2000 CR$_{105}$             & 2008 May 06.989-.999  & 56.57 & 56.46 & 1.02  \nl
                                     & 2008 May 07.000-.066  & 56.57 & 56.46 & 1.02  \nl
                                     & 2008 May 07.972-.999  & 56.57 & 56.48 & 1.02  \nl
                                     & 2008 May 08.000-.060  & 56.57 & 56.48 & 1.02  \nl
                                     & 2009 May 22.963-.999  & 57.08 & 57.22 & 1.01  \nl
                                     & 2009 May 23.000-.013  & 57.08 & 57.22 & 1.01  \nl
(118702) 2000 OM$_{67}$              & 2009 Aug 25.201-.260  & 42.30 & 41.36 & 0.50  \nl
                                     & 2009 Aug 26.178-.231  & 42.30 & 41.36 & 0.48  \nl
(87269) 2000 OO$_{67}$               & 2008 Oct 19.043-.065  & 21.30 & 20.38 & 1.01  \nl   
2000 PE$_{30}$	                     & 2008 May 07.390-.404  & 38.31 & 38.33 & 1.51  \nl
  	                             & 2008 May 08.372-.389  & 38.32 & 38.31 & 1.51  \nl
(82075) 2000 YW$_{134}$              & 2008 May 06.972-.989  & 43.86 & 44.04 & 1.29  \nl   
                                     & 2008 May 07.963-.972  & 43.86 & 44.06 & 1.29  \nl   
(182397) 2001 QW$_{297}$             & 2008 May 07.366-.389  & 47.72 & 48.02 & 1.15  \nl 
                                     & 2008 May 08.357-.372  & 47.72 & 48.00 & 1.16  \nl
                                     & 2008 May 08.390-.404  & 47.72 & 48.00 & 1.16  \nl
2002 GB$_{32}$                       & 2009 May 23.131-.165  & 35.65 & 34.88 & 1.07  \nl   
                                     & 2009 May 24.087-.122  & 35.65 & 34.89 & 1.09  \nl
(84522) 2002 TC$_{302}$              & 2009 Aug 26.278-.306  & 46.59 & 46.13 & 1.11  \nl
(127546) 2002 XU$_{93}$              & 2009 Oct 15.596-.635  & 21.02 & 20.84 & 2.68  \nl
(120132) 2003 FY$_{128}$             & 2008 May 07.214-.233  & 38.46 & 37.65 & 0.90  \nl 
2003 FZ$_{129}$                      & 2008 May 07.106-.174  & 38.19 & 37.29 & 0.70  \nl   
                                     & 2008 May 08.115-.135  & 38.19 & 37.30 & 0.72  \nl   
2003 HB$_{57}$                       & 2009 May 24.122-.156  & 38.10 & 37.26 & 0.87  \nl   
2003 QK$_{91}$                       & 2008 May 07.315-.365  & 39.63 & 39.69 & 1.46  \nl   
                                     & 2008 May 08.339-.357  & 39.63 & 39.68 & 1.46  \nl
2003 UY$_{291}$                      & 2008 Dec 03.154-.228  & 43.29 & 42.32 & 0.19  \nl  
                                     & 2009 Jan 28.000-.104  & 43.31 & 42.90 & 1.19  \nl   
                                     & 2009 Aug 25.383-.421  & 43.38 & 43.41 & 1.33  \nl 
2004 OJ$_{14}$                       & 2009 Aug 25.142-.200  & 45.39 & 44.54 & 0.69  \nl
                                     & 2009 Aug 26.144-.178  & 45.40 & 44.55 & 0.70  \nl
2004 VN$_{112}$                      & 2008 Oct 19.087-.105  & 47.33 & 46.37 & 0.35  \nl   
                                     & 2008 Dec 03.052-.083  & 47.33 & 46.56 & 0.75  \nl   
2004 XR$_{190}$                      & 2008 Dec 03.238-.267  & 58.08 & 57.10 & 0.05  \nl   
2005 EO$_{297}$                      & 2008 May 07.070-.104  & 41.71 & 41.47 & 1.35  \nl   
                                     & 2008 May 08.062-.115  & 41.71 & 41.49 & 1.35  \nl 
2005 PU$_{21}$                       & 2009 Aug 26.101-.143  & 42.84 & 41.87 & 0.37  \nl
2005 SD$_{278}$                      & 2009 Aug 25.354-.383  & 41.75 & 41.53 & 1.36  \nl
                                     & 2009 Aug 26.373-.390  & 41.75 & 41.51 & 1.35  \nl
(145480) 2005 TB$_{190}$             & 2008 May 07.405-.417  & 46.47 & 46.91 & 1.11  \nl 
                                     & 2008 May 08.405-.419  & 46.47 & 46.89 & 1.12  \nl 
2006 SQ$_{372}$                      & 2008 Oct 19.000-.038  & 41.81 & 42.64 & 0.75  \nl    
2007 JJ$_{43 }$                      & 2008 May 07.176-.212  & 41.86 & 40.87 & 0.34  \nl   
2007 TG$_{422}$                      & 2008 Dec 03.229-.238  & 35.71 & 34.85 & 0.79  \nl   
2007 VJ$_{305}$                      & 2009 Aug 25.261-.288  & 35.36 & 34.59 & 1.07  \nl
                                     & 2009 Aug 26.232-.270  & 35.36 & 34.58 & 1.05  \nl
2008 KV$_{42}$                       & 2009 May 23.166-.286  & 31.30 & 30.56 & 1.29  \nl   
                                     & 2009 May 24.168-.224  & 31.29 & 30.56 & 1.29  \nl   
2008 OG$_{19}$                       & 2009 Aug 25.083-.142  & 38.68 & 37.86 & 0.87  \nl
                                     & 2009 Aug 26.085-.101  & 38.68 & 37.87 & 0.89  \nl
2008 YB$_{3}$                        & 2009 May 23.013-.042  &  7.09 &  7.15 & 8.14  \nl  
\enddata
\tablenotetext{}{Quantities are the heliocentric distance ($R$), geocentric distance ($\Delta$) and phase angle ($\alpha$). UT Date shows the year, month, day and time span of the observations.} 
\end{deluxetable}
\end{center}


\newpage



\begin{center}
\begin{deluxetable}{lcccccc}
\tablenum{2}
\tablewidth{6.5 in}
\tablecaption{BVRI Optical Photometry}
\tablecolumns{7}
\tablehead{
\colhead{Name} & \colhead{$m_{R}$} & \colhead{$m_{R}(1,1,0)$} & \colhead{$m_{B}-m_{R}$} & \colhead{$m_{V}-m_{R}$} & \colhead{$m_{R}-m_{I}$} & \colhead{$m_{B}-m_{I}$} \\ \colhead{(mag)} & \colhead{(mag)}  & \colhead{(mag)} & \colhead{(mag)} & \colhead{(mag)} & \colhead{(mag)} & \colhead{(mag)}}  
\startdata
(90377) Sedna            & $20.49\pm0.03$                    &  $1.03\pm0.05$  &  $1.68\pm0.03$ &  $0.61\pm0.03$ &  $0.66\pm0.03$ &  $2.34\pm0.04$ \nl
(48639) 1995 TL$_{8}$    & $21.04\pm0.01$                    &  $4.43\pm0.03$  &  $1.86\pm0.02$ &  $0.68\pm0.02$ &  $0.64\pm0.02$ &  $2.49\pm0.02$ \nl
(19308) 1996 TO$_{66}$   & $21.10\pm0.01$                    &  $4.36\pm0.02$  &  $1.06\pm0.02$  & $0.38\pm0.02$ &  $0.33\pm0.02$ &  $1.38\pm0.02$ \nl
(181874) 1999 HW$_{11}$  & $22.93\pm0.03$                    &  $6.60\pm0.05$  &  $1.32\pm0.03$ &  $0.48\pm0.03$ &  $0.52\pm0.03$ &  $1.84\pm0.04$ \nl
(44594) 1999 OX$_{3}$    & $20.98\pm0.01$                    &  $6.73\pm0.02$  &  $1.82\pm0.02$ &  $0.67\pm0.02$ &  $0.69\pm0.02$ &  $2.52\pm0.02$ \nl
(148209) 2000 CR$_{105}$ & $23.84\pm0.04$                    &  $6.15\pm0.07$  &  $1.26\pm0.09$  & $0.46\pm0.09$ &  $0.59\pm0.09$ &  $1.85\pm0.11$ \nl
(118702) 2000 OM$_{67}$  & $23.29\pm0.03$                    &  $7.00\pm0.05$  &  $1.29\pm0.04$ &  $0.47\pm0.04$ &  $0.59\pm0.04$ &  $1.88\pm0.04$ \nl
(87269) 2000 OO$_{67}$   & $22.11\pm0.02$                    &  $8.76\pm0.04$  &  $1.69\pm0.03$ &  $0.61\pm0.03$ &  $0.61\pm0.03$ &  $2.30\pm0.03$ \nl
2000 PE$_{30}$	         & $21.61\pm0.02$                   &   $5.53\pm0.03$  &  $1.19\pm0.04$  & $0.43\pm0.04$ &  $0.38\pm0.03$ &  $1.58\pm0.04$ \nl
(82075) 2000 YW$_{134 }$ & $20.86\pm0.03$                    &  $4.22\pm0.05$  &  $1.50\pm0.04$ &  $0.55\pm0.04$ &  $0.55\pm0.03$ &  $2.05\pm0.04$ \nl  
(182397) 2001 QW$_{297}$ & $23.46\pm0.05$                    &  $6.47\pm0.08$  &  $1.60\pm0.07$ &  $0.58\pm0.07$ &  $0.67\pm0.06$ &  $2.27\pm0.07$ \nl
2002 GB$_{32 }$          & $23.11\pm0.01$                    &  $7.47\pm0.03$  &  $1.39\pm0.02$ &  $0.51\pm0.02$ &  $0.61\pm0.02$ &  $2.00\pm0.03$ \nl
(84522) 2002 TC$_{302}$  & $20.38\pm0.02$                    &  $3.54\pm0.04$  &  $1.76\pm0.02$ &  $0.64\pm0.02$ &  $0.66\pm0.02$ &  $2.42\pm0.03$ \nl
(127546) 2002 XU$_{93}$  & $21.15\pm0.01$                    &  $7.52\pm0.03$  &  $1.20\pm0.02$ &  $0.44\pm0.02$ &  $0.38\pm0.02$ &  $1.58\pm0.02$ \nl
(120132) 2003 FY$_{128}$ & $20.28\pm0.01$                    &  $4.34\pm0.02$  &  $1.65\pm0.02$ &  $0.60\pm0.02$ &  $0.55\pm0.03$ &  $2.20\pm0.03$ \nl
2003 FZ$_{129}$          & $22.75\pm0.02$                    &  $6.87\pm0.04$  &  $1.32\pm0.04$ &  $0.48\pm0.04$ &  $0.46\pm0.03$ &  $1.78\pm0.04$ \nl
2003 HB$_{57 }$          & $23.15\pm0.02$                   &   $7.25\pm0.02$  &  $1.31\pm0.03$ &  $0.48\pm0.03$ &  $0.54\pm0.03$ &  $1.84\pm0.04$ \nl
2003 QK$_{91}$           & $22.95\pm0.03$                    &  $6.73\pm0.05$  &  $1.37\pm0.04$ &  $0.50\pm0.04$ &  $0.47\pm0.03$ &  $1.84\pm0.04$ \nl
2003 UY$_{291}$          & $23.36\pm0.05$\tablenotemark{a}  &   $5.78\pm0.08$  &  $1.39\pm0.07$ &  $0.51\pm0.07$ &  $0.67\pm0.07$ &  $2.06\pm0.08$ \nl
2004 OJ$_{14}$           & $23.52\pm0.02$                   &   $6.88\pm0.04$  &  $1.42\pm0.03$ &  $0.52\pm0.03$ &  $0.54\pm0.03$ &  $1.96\pm0.04$ \nl
2004 VN$_{112}$          & $22.82\pm0.04$                    &  $5.98\pm0.06$  &  $1.42\pm0.06$ &  $0.52\pm0.06$ &  $0.45\pm0.06$ &  $1.87\pm0.06$ \nl
2004 XR$_{190}$          & $21.54\pm0.03$                   &   $3.93\pm0.05$  &  $1.24\pm0.04$ &  $0.45\pm0.04$ &  $0.52\pm0.04$ &  $1.76\pm0.04$ \nl
2005 EO$_{297}$          & $23.41\pm0.03$\tablenotemark{a}   &  $7.00\pm0.06$  &  $1.32\pm0.05$ &  $0.48\pm0.05$ &  $0.57\pm0.08$ &  $1.89\pm0.08$ \nl
2005 PU$_{21}$           & $22.36\pm0.02$                   &   $6.03\pm0.04$  &  $1.79\pm0.02$ &  $0.65\pm0.02$ &  $0.68\pm0.02$ &  $2.47\pm0.02$ \nl
2005 SD$_{278}$          & $22.11\pm0.02$                   &   $5.70\pm0.04$  &  $1.53\pm0.02$ &  $0.56\pm0.02$ &  $0.53\pm0.02$ &  $2.06\pm0.03$ \nl
(145480) 2005 TB$_{190}$ & $20.86\pm0.02$                   &   $3.99\pm0.04$  &  $1.54\pm0.03$ &  $0.56\pm0.03$ &  $0.55\pm0.03$ &  $2.09\pm0.04$ \nl
2006 SQ$_{372 }$         & $21.48\pm0.02$                   &   $5.11\pm0.04$  &  $1.62\pm0.03$ &  $0.59\pm0.03$ &  $0.65\pm0.04$ &  $2.27\pm0.04$ \nl 
2007 JJ$_{43 }$          & $20.21\pm0.01$                   &   $3.99\pm0.02$  &  $1.61\pm0.02$ &  $0.59\pm0.02$ &  $0.50\pm0.02$ &  $2.12\pm0.02$ \nl
2007 TG$_{422}$          & $21.66\pm0.01$                    &  $6.05\pm0.03$  &  $1.39\pm0.04$ &  $0.51\pm0.04$ &  $0.51\pm0.04$ &  $1.90\pm0.02$ \nl
2007 VJ$_{305}$          & $22.15\pm0.02$                   &   $6.54\pm0.04$  &  $1.44\pm0.03$ &  $0.52\pm0.03$ &  $0.52\pm0.03$ &  $1.96\pm0.03$ \nl
2008 KV$_{42 }$          & $23.47\pm0.04$\tablenotemark{a}   &  $8.36\pm0.07$  &  $1.29\pm0.06$ &  $0.47\pm0.06$ &  $0.42\pm0.06$ &  $1.71\pm0.06$ \nl
2008 OG$_{19}$           & $20.44\pm0.01$                    &  $4.47\pm0.02$  &  $1.47\pm0.01$ &  $0.53\pm0.01$ &  $0.59\pm0.01$ &  $2.06\pm0.01$ \nl
2008 YB$_{3  }$          & $18.24\pm0.01$                    &  $8.41\pm0.02$  &  $1.26\pm0.01$ &  $0.46\pm0.01$ &  $0.49\pm0.01$ &  $1.75\pm0.01$ \nl
\enddata
\tablenotetext{a}{These few objects showed large light variations during the observations indicating possible significant rotational light curves ($> 0.1$ mags).  Their colors were consistent throughout the observations since the variations caused by possible light curves were similar in all filters.  Filters were also rotated after each observation to prevent a light curve from influencing the color calculation. The apparent magnitude ($m_{R}$) and calculated absolute magnitude ($m_{R}(1,1,0)$) are based on the average of the photometry.}
\tablenotetext{}{A few of the above objects have also had colors independently determined.  In most cases the colors reported elsewhere and found in this work are within the uncertainties of the various observations.  1995 TL$_{8}$ has large uncertainties from BVRI data of Doressoundiram et al. (2002) and Delsanti et al. (2001), our results agree with Delsanti et al. (2001) and are inconsistent with Doressoundiram et al. (2002); 1999 HW$_{11}$ has BVR data from Trujillo and Brown (2002); 2000 PE$_{30}$ has BVRI data from Doressoundiram et al. (2001); 2000 YW$_{134}$ has BVRI data from Tegler et al. (2003), Peixinho et al. (2004), Doressoundiram et al. (2007), Jewitt et al. (2007), and Santos-Sanz et al. (2009);  2000 CR$_{105}$ has BVR data from Tegler et al. (2003) and VRI data from Santos-Sanz et al. (2009); 2000 OO$_{67}$ has BVR data with large uncertainties from Tegler et al. (2003); 2003 FY$_{128}$ has VRI data from DeMeo et al. (2009).}
\end{deluxetable}
\end{center}


\newpage



\begin{center}
\begin{deluxetable}{lccccc}
\tablenum{3}
\tablewidth{6.0 in}
\tablecaption{Sloan g',r',i' Optical Photometry}
\tablecolumns{6}
\tablehead{
\colhead{Name} & \colhead{$m_{r'}$} & \colhead{S\tablenotemark{a}} & \colhead{g'-r'} & \colhead{r'-i'} & \colhead{g'-i'} \\ \colhead{} & \colhead{(mag)} & \colhead{} & \colhead{(mag)} & \colhead{(mag)} & \colhead{(mag)} }  
\startdata
(90377) Sedna            & $20.75\pm0.03$                  & $26.3\pm3$  &  $0.85\pm0.03$ &  $0.45\pm0.03$ &  $1.31\pm0.04$ \nl
(48639) 1995 TL$_{8}$    & $21.31\pm0.01$\tablenotemark{b} & $29.4\pm2$  &  $0.96\pm0.02$ &  $0.43\pm0.02$ &  $1.39\pm0.02$ \nl
(19308) 1996 TO$_{66}$   & $21.31\pm0.01$                  & $-0.15\pm1$ &  $0.46\pm0.02$  & $0.12\pm0.02$  & $0.58\pm0.02$ \nl
(181874) 1999 HW$_{11}$  & $23.15\pm0.03$                  & $12.2\pm3$  &  $0.63\pm0.03$ &  $0.31\pm0.03$ &  $0.94\pm0.04$ \nl
(44594) 1999 OX$_{3}$    & $21.25\pm0.01$                  & $31.5\pm2$  &  $0.94\pm0.02$ &  $0.48\pm0.02$ &  $1.42\pm0.02$ \nl
(148209) 2000 CR$_{105}$ & $24.06\pm0.04$                  & $13.7\pm6$  &  $0.59\pm0.09$  & $0.38\pm0.09$  & $0.94\pm0.11$ \nl
(118702) 2000 OM$_{67}$  & $23.52\pm0.03$                  & $14.4\pm3$  &  $0.61\pm0.04$ &  $0.38\pm0.04$  & $0.97\pm0.04$ \nl
(87269) 2000 OO$_{67}$   & $22.37\pm0.02$                  & $24.1\pm3$  &  $0.85\pm0.03$ &  $0.40\pm0.03$ &  $1.26\pm0.03$ \nl
2000 PE$_{30}$	         & $21.83\pm0.02$                  & $4.4\pm3$   &  $0.55\pm0.04$  & $0.17\pm0.03$  & $0.72\pm0.04$ \nl
(82075) 2000 YW$_{134 }$ & $21.10\pm0.03$                  & $17.1\pm3$  &  $0.74\pm0.04$ &  $0.34\pm0.03$ &  $1.05\pm0.04$ \nl  
(182397) 2001 QW$_{297}$ & $23.71\pm0.05$                  & $25.0\pm5$  &  $0.80\pm0.07$ &  $0.46\pm0.06$ &  $1.26\pm0.07$ \nl
2002 GB$_{32 }$          & $23.35\pm0.01$                  & $17.4\pm3$  &  $0.67\pm0.02$ &  $0.40\pm0.02$ &  $1.07\pm0.03$ \nl
(84522) 2002 TC$_{302}$  & $20.65\pm0.02$                  & $28.2\pm3$  &  $0.90\pm0.02$ &  $0.45\pm0.02$ &  $1.34\pm0.03$ \nl 
(127546) 2002 XU$_{93}$  & $21.37\pm0.01$                  & $4.4\pm2$   &  $0.55\pm0.02$ &  $0.17\pm0.02$ &  $0.72\pm0.02$ \nl
(120132) 2003 FY$_{128}$ & $20.54\pm0.01$                  & $20.4\pm3$  &  $0.83\pm0.02$ &  $0.34\pm0.02$ &  $1.17\pm0.02$ \nl
2003 FZ$_{129}$          & $22.98\pm0.02$                  & $9.9\pm3$   &  $0.62\pm0.04$ &  $0.25\pm0.03$ &  $0.88\pm0.04$ \nl
2003 HB$_{57 }$          & $23.38\pm0.02$                  & $12.5\pm3$  &  $0.62\pm0.03$ &  $0.33\pm0.03$ &  $0.95\pm0.04$ \nl
2003 QK$_{91}$           & $23.18\pm0.03$                  & $11.2\pm3$  &  $0.66\pm0.04$ &  $0.26\pm0.03$ &  $0.92\pm0.04$ \nl
2003 UY$_{291}$          & $23.59\pm0.05$\tablenotemark{b} & $19.9\pm7$  &  $0.67\pm0.07$ &  $0.46\pm0.07$ &  $1.14\pm0.08$ \nl
2004 OJ$_{14}$           & $23.76\pm0.02$                  & $15.1\pm3$  &  $0.69\pm0.03$ &  $0.33\pm0.03$ &  $1.02\pm0.04$ \nl
2004 VN$_{112}$          & $23.06\pm0.04$                  & $11.3\pm4$  &  $0.69\pm0.06$ &  $0.24\pm0.06$ &  $1.01\pm0.06$ \nl
2004 XR$_{190}$          & $21.76\pm0.03$                  & $10.3\pm3$  &  $0.58\pm0.04$ &  $0.31\pm0.04$ &  $0.89\pm0.04$ \nl
2005 EO$_{297}$          & $23.64\pm0.03$\tablenotemark{b} & $14.0\pm6$  &  $0.63\pm0.05$ &  $0.36\pm0.08$ &  $0.98\pm0.08$ \nl
2005 PU$_{21}$           & $22.63\pm0.02$                  & $30.0\pm3$  &  $0.92\pm0.02$ &  $0.47\pm0.02$ &  $1.38\pm0.02$ \nl
2005 SD$_{278}$          & $22.36\pm0.02$                  & $17.1\pm3$  &  $0.76\pm0.02$ &  $0.32\pm0.02$ &  $1.09\pm0.03$ \nl
(145480) 2005 TB$_{190}$ & $21.11\pm0.02$                  & $18.1\pm3$  &  $0.76\pm0.03$ &  $0.34\pm0.03$ &  $1.10\pm0.04$ \nl
2006 SQ$_{372 }$         & $21.74\pm0.02$                  & $24.6\pm3$  &  $0.82\pm0.03$ &  $0.44\pm0.04$ &  $1.26\pm0.04$ \nl 
2007 JJ$_{43 }$          & $20.46\pm0.01$                  & $17.7\pm3$  &  $0.81\pm0.02$ &  $0.29\pm0.02$ &  $1.11\pm0.02$ \nl
2007 TG$_{422}$          & $21.89\pm0.01$                  & $13.3\pm1$  &  $0.67\pm0.04$ &  $0.30\pm0.04$ &  $0.98\pm0.02$ \nl
2007 VJ$_{305}$          & $22.39\pm0.02$                  & $14.6\pm3$  &  $0.70\pm0.03$ &  $0.31\pm0.03$ &  $1.01\pm0.03$ \nl
2008 KV$_{42 }$          & $23.70\pm0.04$\tablenotemark{b} & $7.7\pm4$   &  $0.61\pm0.06$ &  $0.21\pm0.06$ &  $0.88\pm0.06$ \nl
2008 OG$_{19}$           & $20.68\pm0.01$\tablenotemark{b} & $18.3\pm2$  &  $0.72\pm0.01$ &  $0.38\pm0.01$ &  $1.10\pm0.01$ \nl
2008 YB$_{3  }$          & $18.46\pm0.01$                  & $9.6\pm0.5$ &  $0.59\pm0.01$ &  $0.28\pm0.01$ &  $0.87\pm0.01$ \nl
\enddata
\tablenotetext{a}{The normalized Spectral gradient for the optical colors of the observed objects (see text for details).}
\tablenotetext{b}{These few objects showed large light variations during the observations indicating possible significant rotational light curves ($> 0.1$ mags).  Their colors were consistent throughout the observations since the variations caused by possible light curves were similar in all filters.  Filters were rotated during the observations in order to prevent any rotational light curve from influencing the color results.  The apparent magnitude is based on the average of the photometry.}
\end{deluxetable}
\end{center}


\newpage

%
%
%
%



\begin{center}
\begin{deluxetable}{lccccc}
\tablenum{4}
\tablewidth{6 in}
\tablecaption{Orbital Information for Observed Objects}
\tablecolumns{6}
\tablehead{
\colhead{Name} & \colhead{Type\tablenotemark{a}}  & \colhead{$q$} & \colhead{$a$} & \colhead{$e$}  & \colhead{$i$} \\ \colhead{} & \colhead{}  & \colhead{(AU)} &\colhead{(AU)} &\colhead{} &\colhead{(deg)} }  
\startdata
(90377) Sedna                        & Oort(3),Det(1,9) &   76.3  & 501   & 0.85  &  11.9   \nl   
(48639) 1995 TL$_{8}$                & Det(9),Obelt(1) &  40.0 & 52.6 & 0.24 & 0.2 \nl
(19308) 1996 TO$_{66}$               & Fam(4) &   38.3  & 43.4  & 0.12  &  27.4   \nl   
(181874) 1999 HW$_{11}$              & Det(1,2) &   39.2  & 52.7  & 0.26  &  17.2   \nl  
(44594) 1999 OX$_{3}$                & Sca(1) &   17.6  & 32.5  & 0.46  &  2.6    \nl   
(148209) 2000 CR$_{105}$             & Det(1,2,9) &   44.1  & 218   & 0.80  &  22.8   \nl 
(118702) 2000 OM$_{67}$              & Det(1) & 39.2 & 100.0 & 0.61 & 23.3 \nl
(87269) 2000 OO$_{67}$               & Oort(5),Sca(1) &   20.8  & 639   & 0.97  &  20.1   \nl   
2000 PE$_{30}$	                     & Det(1) &   35.8  & 54.9  & 0.35  &  18.4   \nl
(82075) 2000 YW$_{134}$              & Det(9),Res(1) &   41.1  & 57.6  & 0.29  &  19.9   \nl   
(182397) 2001 QW$_{297}$             & Det(2),Obelt(1) &   39.7  & 52.0  & 0.24  &  17.0   \nl 
2002 GB$_{32}$                       & Sca(1,2)  &   35.3  & 208   & 0.83  &  14.2   \nl   
(84522) 2002 TC$_{302}$              & Det,Res(1) & 39.2 & 55.3 & 0.29 & 35.0 \nl
(127546) 2002 XU$_{93}$              & High Incl(6) & 21.0 & 66.5 & 0.68 & 78.0 \nl
(120132) 2003 FY$_{128}$             & Det(1),Obelt &   37.1  & 49.3  & 0.25  &  11.8   \nl 
2003 FZ$_{129}$                      & Det(1,9) &   38.0  & 61.7  & 0.38  &  5.8    \nl   
2003 HB$_{57}$                       & Sca(1)  &   38.1  & 160   & 0.76  &  15.5   \nl   
2003 QK$_{91}$                       & Det(1,9) &   38.5  & 68.5  & 0.44  &  4.0    \nl   
2003 UY$_{291}$                      & Det(2,9),Obelt(1) &   41.2  & 49.2  & 0.16  &  3.5    \nl   
2004 OJ$_{14}$                       & Det & 39.4 & 55.7 & 0.29 & 22.4 \nl
2004 VN$_{112}$                      & Det(7) &   47.3  & 337   & 0.86  &  25.6   \nl   
2004 XR$_{190}$                      & Det(8,9),Obelt(1) &   51.3  & 57.4  & 0.11  &  46.7   \nl   
2005 EO$_{297}$                      & Det &   41.1  & 62.4  & 0.34  &  25.1   \nl   
2005 PU$_{21}$                       & Sca & 29.4 & 178 & 0.835 & 6.2 \nl
2005 SD$_{278}$                      & Det & 39.8 & 55.6 & 0.284 & 17.8 \nl
(145480) 2005 TB$_{190}$             & Det(9) &   46.2  & 76.5  & 0.40  &  26.4   \nl 
2006 SQ$_{372}$                      & Oort(5) &   24.2  & 1082  & 0.98  &  19.5   \nl    
2007 JJ$_{43 }$                      & Det,Obelt &   40.2  & 48.3  & 0.17  &  12.0   \nl   
2007 TG$_{422}$                      & Sca       &   35.5  & 528   & 0.93  &  18.6   \nl   
2007 VJ$_{305}$                      & Sca & 35.2 & 199 & 0.823 & 12.0 \nl
2008 KV$_{42}$                       & Retro(6) &   21.2  & 41.8  & 0.49  &  103.4  \nl   
2008 OG$_{19}$                       & Det & 38.6 & 67.4 & 0.428 & 13.1 \nl
2008 YB$_{3}$                        & Retro &   6.5   & 11.6  & 0.44  &  105.1   \nl  
\enddata
\tablenotetext{a}{Det=Detached Disk; Fam= (136108) Haumea (2003 EL$_{61}$) Collisional Family Member; High Incl=High Inclintion Object; Obelt=Outer Classical Belt; Oort=Inner Oort Cloud; Res=Resonace Object; Retro=Retrograde Outer Solar System Object; Sca=Scattered Disk.  References: 1) Gladman et al. 2008; 2) Elliot et al. 2005; 3) Brown et al. 2004; 4) Ragozzine and Brown 2007; 5) Kaib et al. 2009; 6) Gladman et al. 2009; 7) Becker et al. 2008; 8) Allen et al. 2006; 9) Lykawka and Mukai 2007a.}
\tablenotetext{}{Quantities are the perihelion distance ($q$), semi-major axis ($a$), eccentricity ($e$) and inclination ($i$).  Data taken from the Minor Planet Center.} 
\end{deluxetable}
\end{center}


\newpage

%
%
%
%



\begin{center}
\begin{deluxetable}{lcc}
\tablenum{5}
\tablewidth{5 in}
\tablecaption{Spectral Gradients of Small Solar System Objects}
\tablecolumns{3}
\tablehead{
\colhead{Name} & \colhead{S\tablenotemark{a}} & \colhead{Reference} \\ \colhead{} & \colhead{} & \colhead{}}  
\startdata
====  Neutral  Color  ====                 &  ===       &  == \nl
Haumea KBO Family          &   $0.7\pm2$   & 1,9,13 \nl
C-Type Asteroids           &   $2.0\pm2$   & 20 \nl
===Moderate Red Color===      &  ===       & ==  \nl
Dead Comets                &   $7.2\pm5$   & 2 \nl
Jupiter Trojans            &   $7.6\pm3$   & 6,18,19,20 \nl
Irregular Satellites       &   $7.8\pm5$   & 7,10,11 \nl
Neutral Centaur Lobe       &   $8.2\pm5$   & 8,15,17,21 \nl
Outer Retrograde           &   $8.7\pm2$   & 1 \nl
D-Type Asteroids           &   $8.9\pm3$   & 5,20 \nl
Neptune Trojans            &   $9.1\pm3$   & 4 \nl
Comets                     &   $10.0\pm3$   & 2,14,22 \nl
=====    Red  Color    =====                 &   ===      &  == \nl
Scattered KBOs             &  $10.1\pm5$   & 1,8,12  \nl
Damocloids                 &  $11.9\pm3$   & 3 \nl
Detached Disk              &  $14.5\pm5$   & 1 \nl
Plutinos                   &  $16.9\pm5$   & 8,12  \nl
High Incl. Classical       &  $19.9\pm10$  & 8,12  \nl
All Centaurs               &  $20.5\pm15$  & 8,12,15,21  \nl
All KBOs                   &  $20.7\pm15$  & 8,12,15  \nl
All Cubewanos              &  $22.4\pm15$  & 8,12,15  \nl
=== Very Red  Color ===             &   ===      & ==  \nl
Outer Classical Belt       &  $22.8\pm5$   & 1 \nl
High Order Resonance       &  $23.7\pm10$  & 1,8,12 \nl
=== Ultra-Red Color ===               &   ===      & == \nl
Inner Oort Cloud           &  $25.0\pm2$   & 1 \nl
Low Incl. Classical        &  $27.4\pm5$   & 8,12,16  \nl
Red Centaur Lobe           &  $34.3\pm5$   & 8,15,17,21  \nl
\enddata
\tablenotetext{a}{Spectral gradient as defined in the text using known B or g' and I or i'-band photometry normalized to the V-band.  The $\pm$ on the spectral gradient is not an error but displays the general range the type of objects span.}
\tablenotetext{}{References: 1) This Work; 2) Jewitt 2002; 3) Jewitt 2005; 4) Sheppard and Trujillo 2006; 5) Fitzsimmons et al. 1994; 6) Fornasier et al. 2007; 7) Grav et al. 2003; 8) Barucci et al. 2005a; 9) Tegler and Romanishin 2003; 10) Grav et al. 2003; 11) Grav and Bauer 2007; 12) Hainaut and Delsanti 2002 (including updated MBOSS website); 13) Ragozzine and Brown 2007; 14) Lamy and Toth 2009; 15) Peixinho et al. 2004; 16) Peixinho et al. 2008; 17) Peixinho et al. 2003; 18) Karlsson et al. 2009; 19) Melita et al. 2008; 20) Roig et al. 2008; 21) Bauer et al. 2003; 22) Snodgrass et al. 2008.}
\end{deluxetable}
\end{center}


\newpage

%
%
%
%



\begin{center}
\begin{deluxetable}{llccccc}
\tablenum{6}
\tablewidth{6 in}
\tablecaption{T-test and Kolmogorov-Smirnov Test Results}
\tablecolumns{7}
\tablehead{
\colhead{Type 1\tablenotemark{a}} & \colhead{Type 2\tablenotemark{a}} & \colhead{N\tablenotemark{b}}  & \colhead{t-stat\tablenotemark{c}}  & \colhead{t-test\tablenotemark{d}} & \colhead{D-stat\tablenotemark{e}}  & \colhead{K-S\tablenotemark{d}}  \\ \colhead{} & \colhead{} & \colhead{} & \colhead{} & \colhead{} & \colhead{}}  
\startdata
Detached Disk & Cold Classical      & 41 & 8.88  & 99.99\% & 0.88  & 99.99\% \nl
Scattered Strict & Cold Classical   & 35 & 10.49 & 99.98\% & 0.86  & 99.98\% \nl
 $3\sigma$ &                        & -  &  -     &  -       &  -     & 99.73\% \nl
High Order Res. & Cold Classical    & 38 & 2.48  & 83\%    & 0.56  & 98.97\% \nl
Outer Belt & Scattered Strict       & 12 & -3.76 & 99.88\% & 0.86  & 98.96\% \nl
Detached Disk & Outer Belt          & 18 & 3.87  & 97.7\%  & 0.77  & 98.79\% \nl
Detached Disk & Scattered Strict    & 20 & -1.68 & 98.1\%  & 0.64  & 97.3\% \nl
$2\sigma$ &                         & -   &   -    &    -     &   -    & 95\%    \nl
Outer Belt & Cold Classical         & 33 & 3.20  & 92\%    & 0.56  & 92\% \nl
High Order Res. & Scattered Strict  & 17 & -2.08 & 99.09\% & 0.56  & 90\% \nl
Detached Disk & High Order Res.     & 23 & 2.70  & 88\%    & 0.42  & 80\% \nl
Outer Belt & High Order Res.        & 15 & 1.13  & 10\%    & 0.50  & 75\% \nl
\enddata
\tablenotetext{a}{The dynamical groups being tested.}
\tablenotetext{b}{The number of objects used in the test.}
\tablenotetext{c}{The t-statistic from the t-test.}
\tablenotetext{d}{The level of confidence that the two groups are not drawn from the same parent population using the t-test or Kolmogorov-Smirnov (K-S) test.}
\tablenotetext{e}{The D-statistic from the Kolmogorov-Smirnov test.}
\end{deluxetable}
\end{center}


\newpage

%
%
%
%



\begin{center}
\begin{deluxetable}{lcccc}
\tablenum{7}
\tablewidth{5 in}
\tablecaption{Pearson Correlations for Spectral Gradients and Dynamics}
\tablecolumns{5}
\tablehead{
\colhead{Parameter 1\tablenotemark{a}} & \colhead{Parameter 2\tablenotemark{a}} & \colhead{N\tablenotemark{b}}  & \colhead{$r_{corr}$\tablenotemark{c}}  & \colhead{Sig\tablenotemark{d}}  \\ \colhead{} & \colhead{} & \colhead{} & \colhead{} & \colhead{}}  
\startdata
S(Det,Out,ScatStr,LowIncl)  & e & 53 & -0.80 & $99.99\%$ \nl
S(Det,Out,ScatStr)          & e & 25 & -0.53 & $99.1\%$ \nl
S(Det,Out,ScatStr,HighRes)  & e & 35 & -0.41 & $98.5\%$ \nl
S(Det,Out)                     & e & 18 & -0.49 & $97\%$ \nl
S(Det,Out,ScatStr)          & q & 25 & 0.39  & $85\%$ \nl
S(Det,Out)                     & q & 18 & -0.18 & $60\%$\nl
S(Det,Out,ScatStr)          & i & 25 & 0.06  & $20\%$ \nl
S(Det,Out)                     & i & 18 & -0.03 & $0\%$ \nl
\enddata
\tablenotetext{a}{The two parameters being compared through the Pearson correlation coefficient. S is the spectral gradient with the dynamical groups used in () where Det=Detached Disk, Out=Outer main belt, ScatStr=Scattered Strict, LowIncl=Low inclination ``cold'' classical belt and HighRes=High order Neptune resonance objects.}
\tablenotetext{b}{The number of objects used in the correlation.}
\tablenotetext{c}{$r_{corr}$ is the Pearson correlation coefficient..}
\tablenotetext{d}{Sig is the confidence of significance of the correlation.}
\end{deluxetable}
\end{center}


\newpage

\end{document}